# The physics, travels, and tribulations of Ronald Wilfrid Gurney

Brian Pollard[a] and Saman Alavi[b]

[a] H. H. Wills Physics Laboratory, University of Bristol

[b] Department of Chemistry and Biomolecular Sciences, University of Ottawa

## Abstract

Ronald Wilfrid Gurney is one of the lesser-known research students of the Cavendish Laboratory in the mid 1920s. Gurney made significant contributions to the application of quantum mechanics to problems related to tunneling of alpha-particles from nuclei, to formation of images in photographic plates, the understanding of the origin of color-centres in salt crystals, and in the theory of semiconductors. He was the first physicist to apply quantum mechanics to the theory of electrochemistry and ionic solutions. He also made fundamental contributions to ballistics research. Gurney wrote a number of textbooks on fundamental and applied quantum mechanics in a distinctive style which are still useful as educational resources. In addition to his scientific contributions, he travelled extensively, and during and after World War II worked in the United States. During the cold war, he got entangled in the Klaus Fuchs affair and lost his employment. He died at the age of 54 in 1953 from a stroke. With the approach of the 100 year anniversary of quantum mechanics, it is timely to commemorate the life and contributions of this somewhat forgotten physicist.

## Introduction

As the centenary of quantum mechanics approaches, it is timely to re-explore the life and contributions of Ronald Wilfrid Gurney (1898-1953), a somewhat forgotten physicist. Gurney's life captures the zeitgeist of Britain and the United States in the first half of the twentieth century. He lost a brother in World War I and served in the army, earned a doctorate in Cambridge under Ernest Rutherford, and travelled widely in the interwar period, including to Europe, the United States, Japan, India, and the Soviet Union. He made fundamental contributions to quantum mechanical understanding of alpha-particle tunnelling and nuclear structure in Princeton, was involved in applying quantum mechanical concepts to the physics of solids and solution physical chemistry emphasizing electron transfer and ionic processes, and discovered the photophysics of photographic latent image formation while in Bristol, and made fundamental contributions to ballistics research at the Aberdeen Proving Grounds during World War II. Gurney wrote five textbooks which helped disseminate quantum mechanical ideas to non-theorists and are exemplary in their clarity. His writing highlights physical intuition, without avoiding mathematical technicalities, and even now his books are valuable reading as supplementary sources. Gurney became entangled in the Klaus Fuchs affair while at Columbia University, Johns Hopkins University, and the University of Maryland through indirect connections and politics of the McCarthy era. He suffered loss of work and isolation and died at the young age of 54 from a stroke.

Gurney had personal contact with many prominent physicist and chemists in the first half of the twentieth century, including Ernest Rutherford, Ralph H. Fowler, Nevill F. Mott, Edward U.

Condon, J. Robert Oppenheimer, Joseph E. Mayer, Maria Goeppert Mayer, Isidor I. Rabi, Louis P. Hammett, Edward Teller, Llewellyn H. Thomas, Klaus Fuchs, and many others.

Two prominent physicists Edward U. Condon [1,2] and Sir Nevill F. Mott [3] wrote tributes to Gurney's scientific contributions and personality. In his obituary Sir Nevill Mott, a friend and one-time collaborator wrote: "It was typical of him that, though he could use mathematics as well as another, he never trusted it; no theory was a theory to him until he could grasp it intuitively and, above all, draw diagrams of it. In this he was among theorists perhaps the truest representative of Rutherford's school, and of the British tradition. Wherever Gurney went - and not least at Bristol - he contributed to the work of the place a capacity to see things clearly, graphically, and in their essentials. He was a man with whom it was a privilege to work, and from whose unique qualities of mind and character his collaborators could learn much." [3]

Condon's biography paints a vivid image of Ronald Gurney and his wife Natalie and was a motivation for us to learn more about the Gurneys. Memories of Ronald Gurney's life, work, and tribulations have been mostly forgotten and with the centenary of the discovery of quantum mechanics upon us, it timely to revisit the life and contributions of this brilliant and underappreciated physicist from the early days of quantum mechanics.

**Childhood and Cambridge**
Ronald W. Gurney was born on July 1, 1898, in Cheltenham, England, the son of Emily Constance Gurney (née Taunton, 1858-1949) and Walter Gerald Gurney (1859-1930). He was the youngest child of the family with siblings Kenneth Gerard (1888-1917), Gwynneth Constance (1890-1976), and Brian Taunton (1894-1965). His father was a solicitor in the legal firm Winterbotham, Gurney and Co. and an alderman in the Cheltenham town council.[4] The Gurneys were Nonconformists and this seems to have influenced his sphere of friends and social circle, even later in life. His eldest brother Kenneth was a Oxford-trained lawyer, His sister Gwynneth never married, was a long-time teacher in a girl's grammar school, and lived in the family home in Cheltenham her entire life. Ronald's brother Brian became a solicitor and took over the family legal firm after his father's death.[4] None of the siblings had children and it has become difficult to trace personal stories and photographs from Ronald's youth. The only known scientific connection in the Gurney family was that Ronald's cousin Janet Viney was the mother of Paul Taunton Matthews FRS.[5] The family was wealthy and this may also have been a contributing factor influencing Gurney's proclivities as life and career choices unfolded.

Ronald Gurney entered Cheltenham College, a private school in Gloucestershire, in September 1911 and left in April 1917. [6-8] Ronald's brother Kenneth fought in the British Army in World War I. He was wounded during the Battle of Cambrai in early December 1917, was taken by German soldiers to the hospital, and died from his wounds as a prisoner of war on the 17th December, 1917.[9,10] Ronald was called into the military in 1918 and received his commission as "ensign" (2nd Lieutenant) in the 4th (City of Bristol) Battalion, the Gloucestershire Regiment on 3rd February, 1919.[11,12] Due to his age, he avoided military action in World War I.

After release from the army, Ronald matriculated in Trinity Hall, Cambridge University in physics, chemistry, and geology on October 1919, see Fig. 1. He graduated with a B.A., Hons. in June 1922. [12,13]

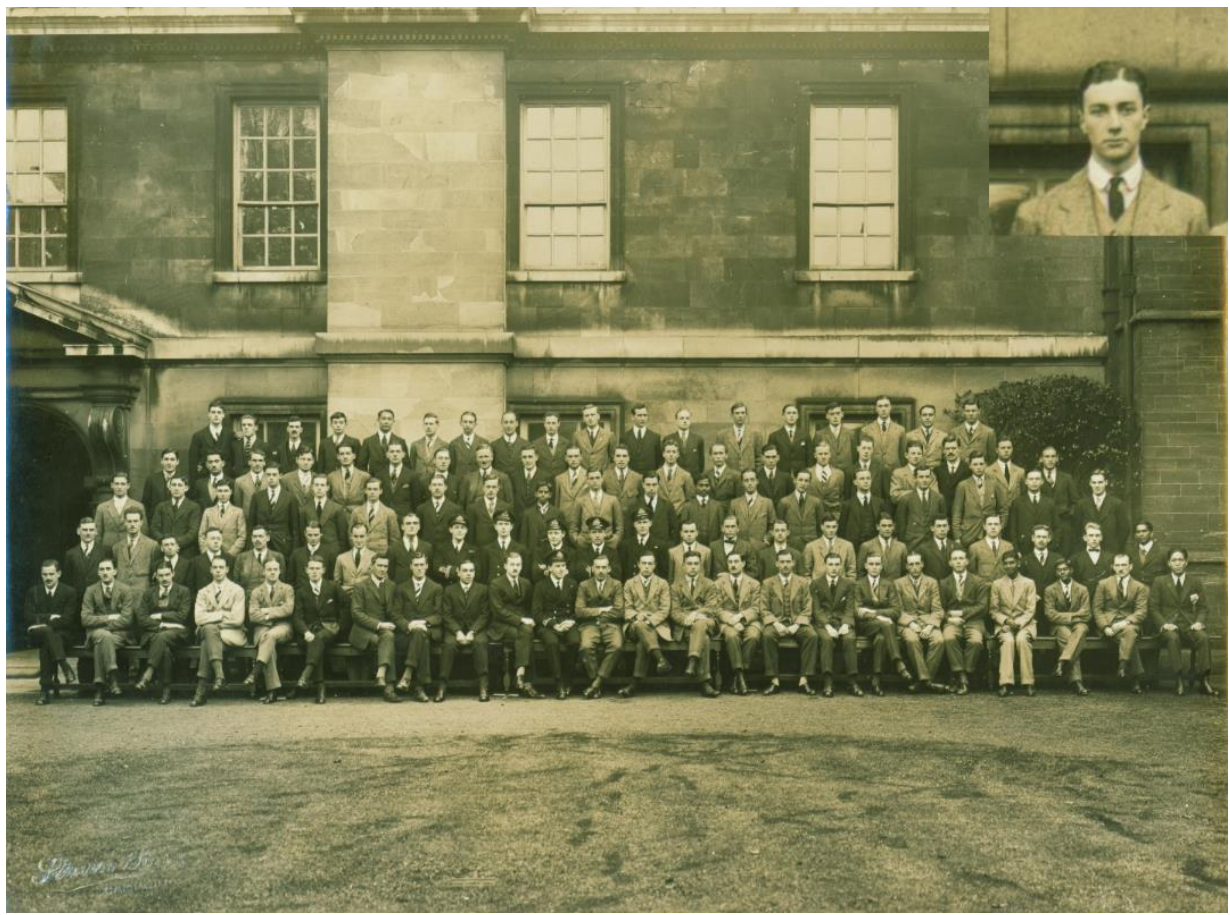

**Figure 1**. Matriculation photo of the 1919 class at Trinity Hall, Cambridge University. Gurney is third from the right on the top row. A close-up of Gurney is shown in the inset. [*Published courtesy of Trinity Hall*]

Gurney started graduate studies in experimental physics under Sir Ernest Rutherford at the Cavendish Laboratory. With the presence of many prominent researchers (some of whom are listed in the caption of Figure 2), the Cavendish was a world-leading centre of experimental physics. Fellow graduate students working at the Cavendish at the time included Charles Eryl Wynn-Williams, Cecil Powell, Herbert Skinner, and J. Robert Oppenheimer, with whom Gurney and his future wife established friendship. Gurney studied in Cambridge at the same time as Paul Dirac, but there is no record of them interacting during this time. Gurney taught physics, mechanics, elementary mathematics, and radiology while in Cambridge.

Gurney received an M.A. degree and published two research papers in 1925 which studied the ionization of monatomic and diatomic gases irradiated by alpha-particles.(1,2)[*] His Cambridge papers were published in the *Proceedings of the Royal Society A* (PRSA) and were communicated by Sir Ernest Rutherford.

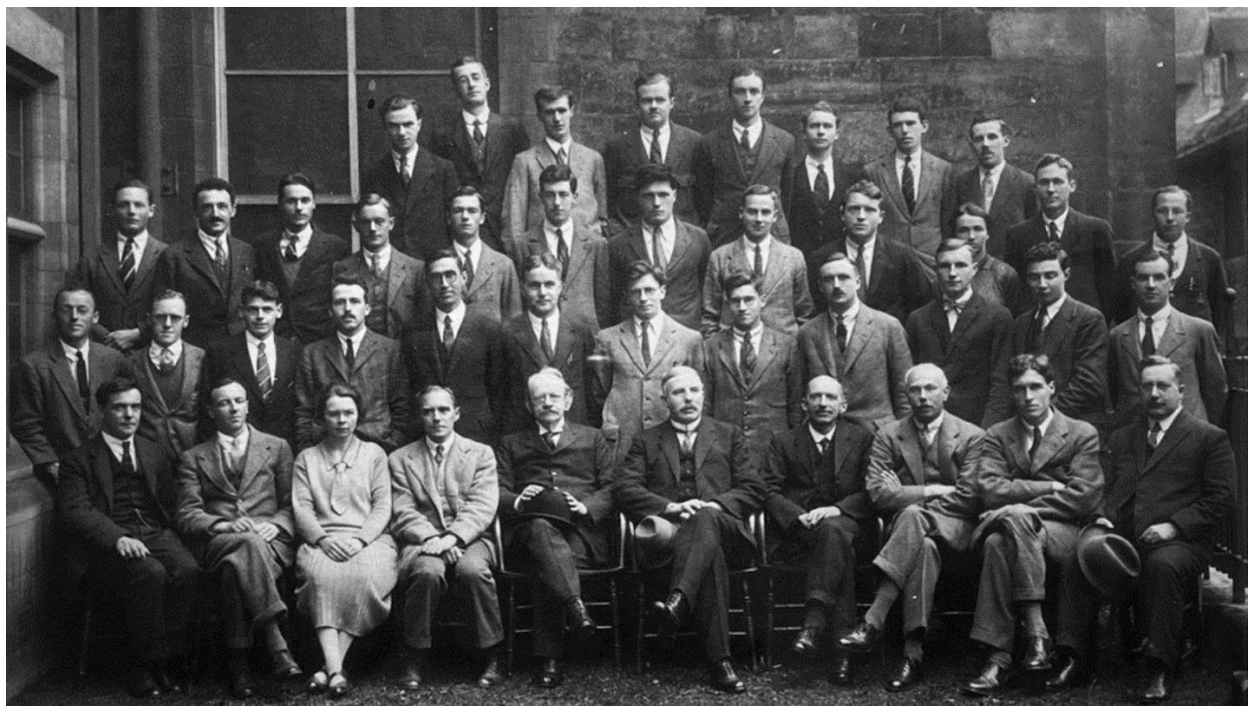

**Figure 2.** Cavendish Laboratory staff and research students in 1926 showing many prominent figures. First row from left to right: P. L. Kapitsa, J. Chadwick, K. B. Blodgett, G. I. Taylor, J. J. Thompson, E. Rutherford, C. T. R. Wilson, F. W. Aston, P. M. S. Blackett, and Chamberlain. Gurney is fifth from the left on the top row. J. R. Oppenheimer is second from the right in the 2nd row, C. Powell is fourth from the left in the 3rd row, C. E. Wynn-Williams is first from left and H. Skinner is third from left in the top row. [*Courtesy of the Cavendish Laboratory, University of Cambridge*]

In his next two papers in 1926, Gurney used magnetic deflection to study the energy spectrum (velocity distribution) of beta-particles released from the radium decay products Radium B and Radium C,(3) and thorium decay products Thorium B and Thorium (C+D). (4) As shown in Fig. 3, the beta-particle energy spectrum is verified as being a superposition of discrete lines superimposed on a continuous background. These two papers and knowledge of Geiger's results from 1921 regarding the energy spectra of emitted alpha-particles,[14,15] gave Gurney intuition on the presence of discrete energy states in the nuclear structure, a point he later used in analysing tunneling and adsorption of alpha-particles in nuclei. In both papers, Gurney acknowledges the assistance of Dr. Charles D. Ellis. Gurney's experiments where perhaps designed in the context of the scientific controversy between Lise Meitner and Ellis on the interpretation of the sources of the spectral lines and continuous background of the beta-particle energy spectrum.[16]

---

[*] Citations to Gurney's publications are enclosed in parentheses and are given in a separate section at the end of the text.

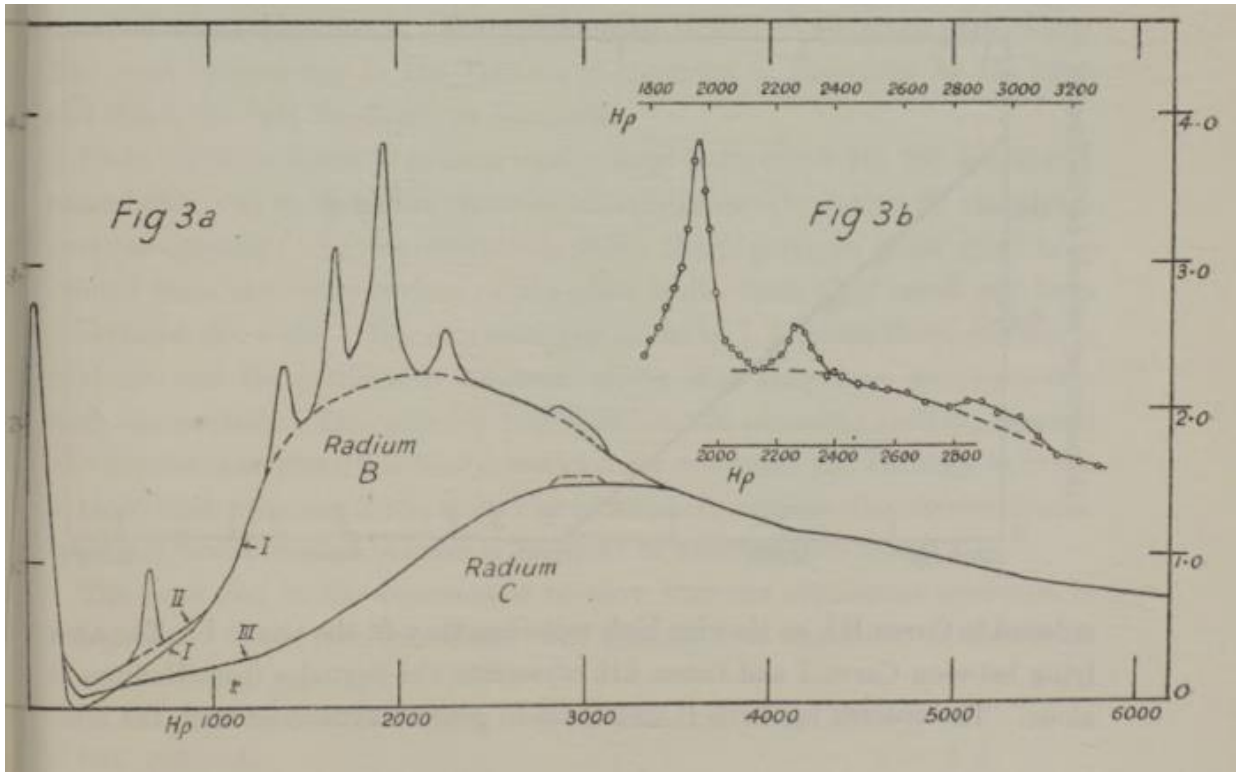

**Figure 3.** The continuous beta-particle spectrum from Radium B and C with the superimposed discrete spectral lines. (3) Studying these spectra may have led Gurney to presence of discrete resonance energies in the nuclear structure which he later used in his studies of tunnelling and resonance adsorption of particles by nuclei. (13)

Gurney graduated with a Ph.D. degree in June 1926 and with a recommendation from Rutherford, was granted a one-year International Education Board and Commonwealth Fund Fellowship from the Rockefeller Foundation to study at Princeton University.[17]

Sources on both sides of the political spectrum have claimed that in the 1920s Ronald Gurney became a member of the Great Britain Communist Party (GBCP)[18] or was a Soviet sympathizer. However, lists of prominent members of the GBCP and published MI5 and FBI files on Gurney do not show evidence or even include accusations of this point, which would have been highly relevant to their investigations.[19,20] His politics leaned left, however, there is no evidence of explicit political activity during his Cambridge period.

**Princeton and International Travels (1926-1931)**

Gurney travelled widely prior to Princeton and had visited Belgium, France, Germany, Austria, Switzerland, and Italy. He could converse in Italian and French and had studied German.[12] Gurney departed England in September 1926 to work at the Palmer Laboratory in Princeton University. [1,17,21] His initial goal was to conduct experimental work with Karl T. Compton (1887-1954), an expert on electronic processes and spectroscopy. His work during the first year at Princeton was sufficiently remarkable that based on the Compton's recommendation, Gurney was granted a Jane Eliza Proctor visiting fellowship and his stay extended to the academic year of 1928.[12,13,17]

Gurney's experimental work in Princeton resulted in two papers (6), (9) submitted in autumn 1928. The first paper studies the scattering of emitted high momentum $K^+$, $Cs^+$, and $Li^+$ ions by a heated platinum surface. (6) He determined the velocity and angular distribution of the ions as a function of the Pt foil surface temperature. The follow-up work, (9) studied the ionization efficiency of hydrogen gas bombarded by 7000 V energy $K^+$ ions. Gurney published a third paper with Philip Morse in 1929 where they analyzed modifications required in positive-particle ray experiments taking into account the electric fields formed by secondary ionization products. (10)

In Princeton, Gurney became interested in the physics of the solar chromosphere, an area were R. H. Fowler was working on in Cambridge.[22] He published a paper in December 1928 which analyzed the practice of determining the boundaries of the solar chromosphere and prominences using calcium atom fluorescent emission lines.(7) A second paper on solar physics from 1928 analyzes the balance of radiation pressure and gravity in maintaining the mass flow of ions from the upper to lower solar chromosphere. (8)

The breakthrough most associated with Gurney today occurred in 1928. According to E. U. Condon (1902 - 1974),[2] Gurney read papers by J. Robert Oppenheimer (January 1928) and Fowler and L. Nordheim (May 1928) regarding the leakage of electrons from atoms or metals through the action of intense electric fields by the mechanism of quantum mechanical tunnelling.[23-25] Gurney had the intuition that a similar tunnelling mechanism can explain alpha-particle decay through the barrier surrounding radioactive atomic nuclei. Lacking theoretical training in quantum mechanics to calculate the consequences of this intuition, he first shared the idea with Princeton mathematical physicist Howard P. Robertson who dismissed it as having no promise. Luckily for Gurney, in the summer of 1928, Condon, a young theoretical physicist with expertise in quantum mechanics, joined the Princeton faculty. Gurney shared his idea with Condon who was enthusiastic.[1] In a short time, the two quantified the argument and submitted a letter to *Nature* (on July 30) outlining their ideas which was published on September 22. (5) Very soon after, perhaps with a sense of urgency, George Gamow, a post-doctoral fellow at the Institute for Theoretical Physics in Copenhagen, described his calculations along the same lines in a letter to *Nature* (submitted on Sept. 29) which was published on November 24.[26] Gamow had earlier submitted detailed calculations on this problem to the *Zeitschrift für Physik* (submitted July 29, received August 1928) which was in press, but not published, at the time.[27] Gurney and Condon submitted a full version of their calculations to the *Physical Review* in November 1928, which was published in February 1929.(11)

The Gurney – Condon model considers the alpha-particle to be imbedded in the potential well of the remainder of the nucleus, as shown in Fig. 4. Outside the nucleus, the potential corresponds to a Coulombic potential for a charge $Z - 2$, where $Z$ is the nuclear charge. Gurney and Condon calculated the leakage factor for alpha-particles which lacked sufficient kinetic energy to classically overcome the peak barrier of the nuclear potential energy curve. The model addressed the stochastic nature of the alpha-particle decay and predicted that the first-order alpha-particle decay rate is independent of the prior history of the nucleus. The model also indicated how slight changes in the kinetic energy of the released alpha-particles lead to very large differences, as high as a factor of $10^{23}$, in the rate of alpha-particle decay for different nuclei. Their calculations reproduced the Geiger – Nuttall laws which relate the life time of a radioactive nucleus to the kinetic energy of the alpha-particles which it emits.(11)

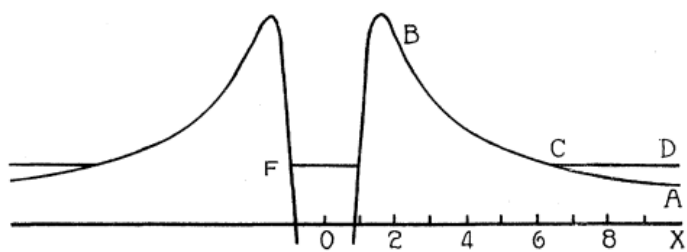

The unit of abscissas is $10^{-12}$ cm. The horizontal line gives the energy of the α-particle emitted by uranium, $6.5 \times 10^{-6}$ ergs.

**Fig. 4**. The potential energy profile for an alpha-particle in the nucleus of thorium $(Z = 90)$ showing Coulombic decay outside of the confines of the nucleus. [*Reproduced with permission from American Physical Society*]

The Gamow and Gurney-Condon models were the first successful application of quantum mechanics to the atomic nucleus and in 1929 generated a great deal of activity.[2] The work excited Bohr and contributed to making Copenhagen a center for theoretical nuclear structure studies.[28] Rutherford was enthusiastic about these theoretical developments and their discussion was an important part of a meeting in Cambridge he hosted on the structure of atomic nuclei. Gamow attended this meeting and Rutherford discussed the implications of the tunnelling in detail and acknowledged the work of Gurney and Condon.[29]

Gurney realized that the tunneling model also predicts a high probability of alpha-particle or proton penetration into the nucleus by those positively charged particles which had energies very near the quasi-stable resonance levels inside the nucleus. This particle penetration could lead to artificial disintegration of the recipient nucleus, a process recently discovered by Patrick M. S. Blackett and Rutherford. Gurney discussed calculating the enhanced penetration probabilities with Condon, but to Condon's later regret, he had misgivings regarding the narrowness of the nuclear resonance energy levels in uranium and radium and recommended against pursuing the idea. Gurney later published a qualitative account of this idea. (13)

Condon writes that prior to his interest in tunnelling, Gurney had a passing familiarity with quantum mechanics and expressed a lack of confidence in his understanding of it.[1] The initial lack of emphasis on the new quantum mechanics in Cambridge in the late 1920's, with the notable exception of Fowler, has been noted in the context of Dirac's early work.[30] Gurney's lack of background in quantitative quantum mechanics, Condon's perceived lack of further interest in pursuing this topic, and the ending of his fellowship at Princeton, on the one hand, and the rapid progress of work on nuclear structure in Copenhagen led by Gamow and Bohr, on the other, resulted in a lost opportunity for Gurney to expand his research in this area and to gain a greater academic standing. In Spring of 1928, at the suggestion of the Director of Personnel at Princeton, Gurney applied for a position at the Physics Department at Antioch College, but nothing came of this.[12] Karl Compton's view of Gurney (expressed in the Proctor Fellowship application) was that Gurney was "not a brilliant physicist, nor a particularly skillful experimenter, but a clear thinker, industrious and able to do good work."[12] Compton ultimately agreed he had done a good project, but did not recommend Gurney for a further renewal of the Proctor scholarship beyond 1928. This was rather unfortunate since at the time of the scholarship renewal, Compton would not have been aware of Gurney's alpha-particle tunnelling work.

From Condon and archives held at Princeton University, we hear of Gurney's extreme shyness.[2,12] The assessment of A. F. West, the Dean of the Graduate College at Princeton was that Gurney was "a rather self-contained man of high ability, who will do his best work in his own way" and "His characteristic English reserve … makes it less easy to form early reliable judgments about the quality of his work".[12] W. F. Magie, Chairman of the Princeton Department of Physics wrote that Gurney "is extremely reticent and it is hard to get from him any knowledge of the work that he is doing".

At the same time, Condon notes occasions where Gurney came out of his shell. Gurney was an accomplished flautist and pianist and a great lover of music. He later composed chamber music pieces which he and friends performed. He was an enthusiastic English country dancer and instructed evening dance classes for faculty couples at Princeton. As early as December 1926, a misunderstanding of the nature of his dance instruction nearly got him into trouble with his scholarship. Word reached A. Trowbridge, director of the International Education Board that Gurney was spending too much time on dance instruction and was not devoted to his studies. However, West vouched that he was only giving unpaid lessons on history and technique of English Folk Dancing once a week and the matter was resolved.[12]

After initiation into quantum mechanics at Princeton, a career pathway for a young researcher in Gurney's position would appear to be a move back to Europe to work at one of the centres of excellence in experimental or theoretical quantum mechanics. However, in January 1929 Gurney left Princeton on a Japanese research fellowship to work at the Institute of Physical and Chemical Research (now RIKEN) in Tokyo.[2,13,31] He initially intended to stay in Japan for six months and return to England afterwards.[12] A disposition towards travel and relative wealth of his family may have blunted some of the urgency of finding a permanent position.

On his way to Japan, Gurney stopped over at the Mount Wilson Observatory, which he gives as his affiliation on a Letter to the Editor in *Nature* published in February 1929. (12) This note is related to spectroscopic measurements of two Fraunhofer lines in the chromosphere spectrum during a solar eclipse. (12) Gurney verified that the disappearance of calcium H and K lines of the spectra at the edge of solar prominences during a solar eclipse can correctly be used as a measure of the farthest extent of the chromosphere. This short paper generated two response letters in *Nature*.[32,33]

There is no record of the research Gurney conducted in Tokyo or of his collaborators there. In an interesting anecdote, while in Japan, Gurney seems to have learned a few Japanese songs which at least on one occasion, he sung with piano accompaniment.[34]

Soon after arriving in Tokyo, Gurney wrote a Letter to *Nature*, entitled Nuclear Levels and Artificial Disintegration (dated Feb. 20, and published April 13, 1929) where he outlined the idea of resonance enhanced absorption of protons by the nucleus. (13) This prediction had not been made by Gamow and presaged experiments on artificial nuclear disintegration of Cockroft and Walton by three years. Condon, William Fowler, and Gamow note that this paper was the first to discuss the possibility of resonance absorption phenomenon in nuclei.[2,35,36] This discovery is usually attributed to the experimental work of the Fermi group in Rome [37] and the theoretical work of Breit and Wigner[38] conducted in 1933-1936. Gurney was not in a position to exploit this important insight in the context of nuclear physics but later applied the idea of resonance tunnelling to his pioneering studies of electron transfer between the electrode and solution in electrochemical processes.

After Japan, in another unorthodox career move, Gurney spent more than a year in India staying with a cousin, Angus Hugh Gurney (1889-1939), who worked as the deputy accountant general in British Punjab.[21,39] Not much is on record of his time in India, other than he seems to have lived well and used this time to study quantum physics. The only mention of Gurney's time in India in a scientific context is a rather curious short letter to *Nature* about an optical illusion related to binocular vision published in 1938 after he had returned to Britian.(24)

Walter Gurney, Ronald's father, died in February 1930.[40] This may have prompted his eventual return to England.

**Return to England, Natalie (1931-1939)**
After a five-year long grand tour, Gurney duly returned to Trinity Hall in 1931. By July, he submitted the first of his papers on electrochemistry entitled "The Quantum Mechanics of Electrolysis".(14) In this paper, in unassuming fashion, Gurney assembled many of the key ideas of modern electron transfer theory in the context of electrochemistry.[6,41] He introduced quantum mechanical metal band theory and the Fermi-Dirac distribution of electrons into the description of electrodes. He recognized the importance of tunnelling in the electron transfer process between the metallic and ionic orbitals near the electrode – solution interface and described the effect of the water solvent on the energy levels of the ions in solution.[6] Gurney used these factors to explain the overpotential required for electrolysis processes.(14) This work seems to have been motivated by discussions with F. P. Bowden in Cambridge who explained the results of his electrochemical experiments to Gurney. Based on this paper, Bockris, Reddy, and Gamboa-Aldeco called Gurney "the first physical electrochemist". [42]

Gurney soon reintegrated into the scientific life at Cambridge. He commented in a *Faraday General Discussion* on The Adsorption of Gases on Solids.[43] In late 1931, Gurney was given a temporary appointment as a Lecturer at Manchester University in the lab of Lawrence Bragg. In January 1932, Gurney published a second paper on the quantum mechanics of electrochemistry,(15) where he used quantum mechanical arguments to demonstrate that the heat evolved during the chemical reaction involving two metals in solution, such as in the reaction $CuSO_4(aq) + Zn(s) \rightarrow ZnSO_4(aq) + Cu(s)$, is related to the Voltaic potential difference between the two metals (Cu and Zn, in this case) in an electrochemical cell. He described the potential energy of metal ions embedded in the metal electrode and hydrated in the adjacent solution as a double well with relative minima determined by the relation between the electrode potential and the concentration (activity) of the ion in the solution. Fowler wrote a statistical mechanical Appendix to this paper which put Gurney's results on a more formal footing.[44]

Gurney submitted his first paper on the physics of ionic solids in February 1933. He used quantum mechanical electron transfer ideas to explain how photoelectric absorption leads to the formation of latent photographic images in halide crystals (16). Gurney argued that energy from light can transfer an electron from a halide anion to a metal cation which resides near the sight of an imperfection (hole) in the crystal lattice. The photoelectron reduces the cation to a metal atom which causes coloration in the halide crystal. The key factor in photo-absorption and image formation is the presence of imperfections in the lattice which provide the crystal with electronic levels that lie between the localized occupied electronic levels of the ions and the high energy unoccupied empty orbitals of the insulating crystal. Gurney and Nevill Mott later modified the ideas of this paper, as described below, to give the correct mechanism of latent image formation.

In Manchester, Gurney did not publish any work with Bragg. He proposed to Bragg that he should work on the mechanics of photoelectric conductivity in cotton and Bragg referred him to the president of the Cotton Industry Research Association of Britain for follow-up, but no record of work in this area is available.[45]

Gurney was soon travelling again. In the summer of 1932 Peter Kapitsa made arrangements for a distinguished party, including his wife, W. L. Bragg and his wife Alice, R. H. Fowler, Paul Dirac, and Gurney to participate in the yearly conference held in Abram F. Ioffe's lab in Leningrad.[46] After obtaining necessary visas, the party travelled by ship to Leningrad. During their voyage, Soviet authorities decided that foreign guests would no longer be permitted to attend Ioffe's annual conferences.[46] However, as this decision was made after the group's departure from London, they were allowed to attend the conference on an unofficial basis. Official "guides" were not assigned to the guests, which gave them the benefit of being able to roam freely in Leningrad without a minder. Bragg reported that Gurney got lost numerous times while wandering in the city.[46] Accounts of Dirac's [30,47,48] and Fowler's[49,50] scientific participation at this conference are available but nothing is reported on Gurney's participation.

The physicist Arthur M. Tyndall of the University of Bristol hired the young theoretical physicist Nevill F. Mott as the Melville Wills Professor in Theoretical Physics in the autumn of 1933. In the 1933-1934 year, Gurney was also hired at the University of Bristol on a George Wills Research Associateship,[51] ostensibly as an experimental appointment. Mott was a Cambridge graduate and friend of Gamow from their time in Copenhagen and would have been familiar with Gurney's work on alpha-particle tunnelling.[52] In Cambridge, Mott worked on quantum mechanical collision theory, and was expected to start a nuclear physics group in Bristol, but his interest in metal / semi-conductor physics became so successful that it became main research theme at the Physics Department in Bristol. The presence of Mott, Herbert Skinner, Harry Jones, and Cecil Powell, and recent refugees from Nazi Germany Walter Heitler, Herbert Fröhlich, and very briefly, Hans Bethe, made this an opportune time to be in Bristol, which became a hub for solid-state physics research. The physics department started a series of specialized international conferences on Physics of Solids, the first of which in 1935 was on Metals and attracted many distinguished visitors (Fig. 5).[53]

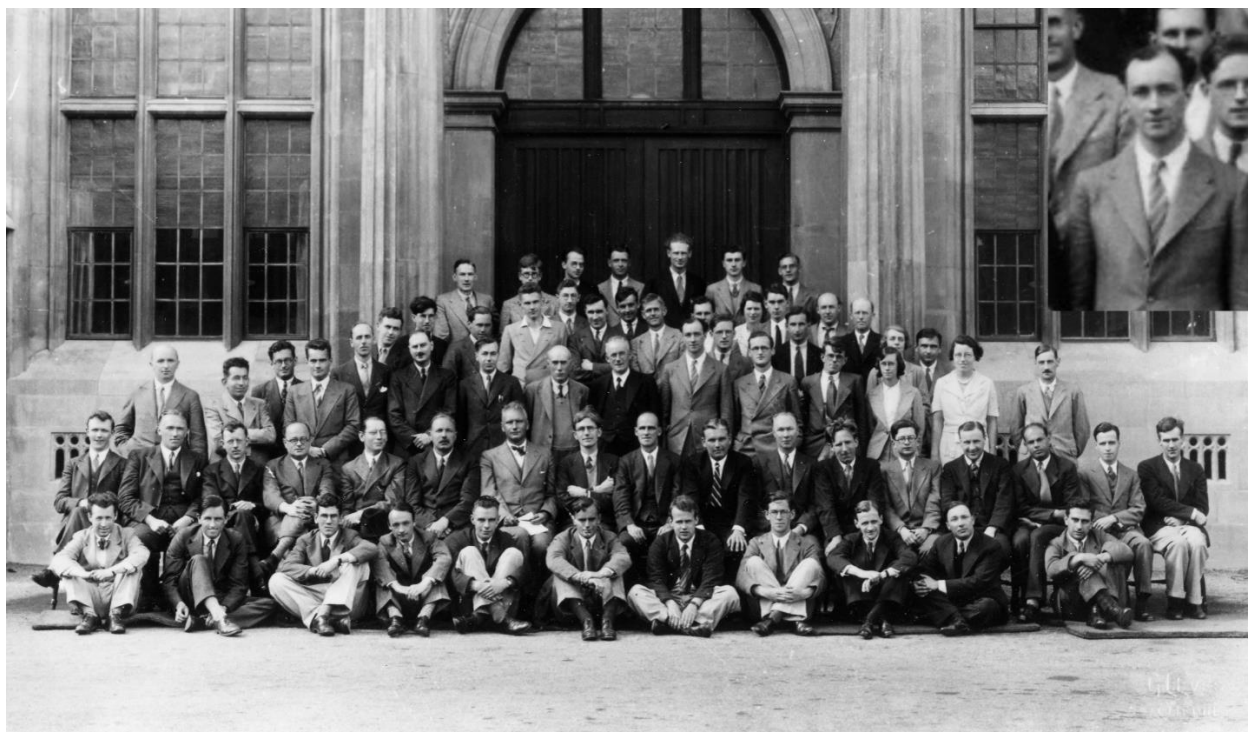

**Figure 5.** The Bristol Metal Conference 1935 in which many prominent physicists including W. Heitler, C. Powell, E. Teller, F. Simon, C. G. Darwin, W. Gerlach, N. F. Mott, A. M. Tyndall, H. F. Mark, J. D. Bernal, R. E. Peierls, H. W. B. Skinner, F. London, H. Fröhlich, R. Smoluchowski, K. Fuchs as participants. Gurney is in the third row, sixth from the right. A close-up of Gurney is shown in the inset at the upper right hand corner. [*Archives of HH Wills Physics Laboratory, University of Bristol, courtesy of AIP Emilio Segrè Visual Archives*]

A major change occurred in Gurney's life when he met Natalie (Nalinka, Natalia) Kouteinikoff (January (or October)[54] 30, 1908 - September 20, 1993), Fig. 6. Natalie was a resident of London, born in St. Petersburg to a Russian physician father and a British mother.[55] Natalie left Russia by the age of four[56] and between the ages of 11 to 16, studied in a boarding school in Eastbourne in Sussex.[57] Natalie was extroverted, warm-hearted, socially engaged, and an avid letter-writer who at times could be quite forthright. Her letters are a main source of information on Ronald Gurney's life. She was a member of the London Young Friends (Quaker) group prior to her marriage. [58,21] This group, in coordination with the Student Christian Movement helped Far Eastern students settle into London.[58] Between 1932 to 1934, she was employed as Secretary of the "German

Refugees Hospitality Committee" formed by the Labor Party in collaborations with Quakers.[20] Natalie and Ronald met in a session of the Progressive League of H. G. Wells, Cyril Joad, and Aldous Huxley[59,60] and the two married on 31st August 1934 in London.[61] Natalie joined Ronald in Bristol, and for the academic year 1935-36 studied at the University of Bristol for qualification as a Health Visitor, but does not seem to have completed the course.[54,62]

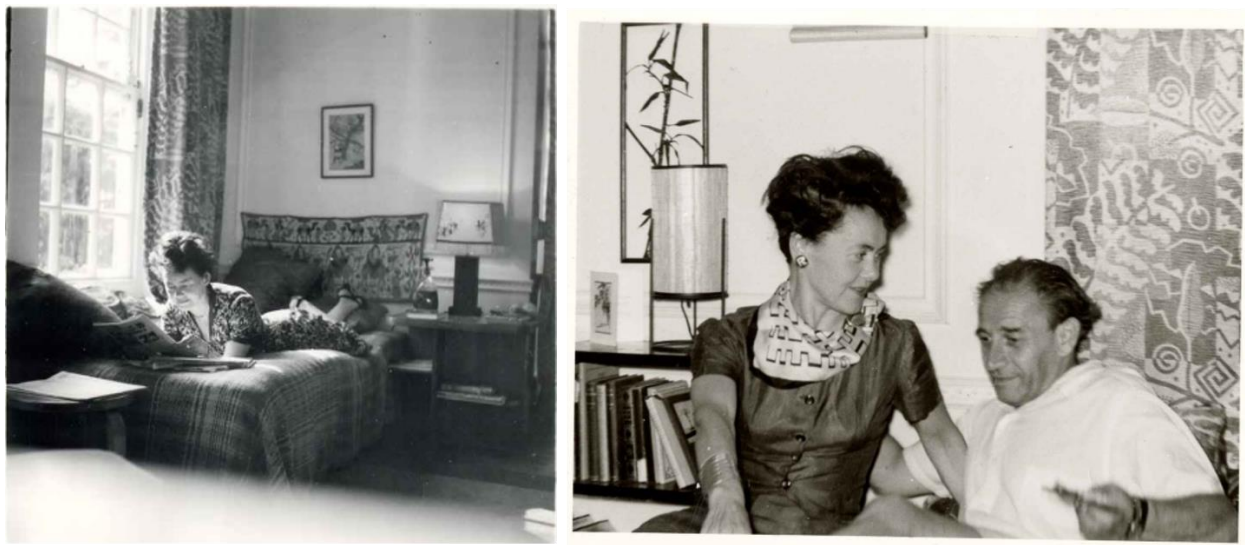

**Figure 6**. (left) Natalie Gurney in her home in Hampstead in 1955. (right) Natalie Gurney with her second husband Fredrick Kräupl Taylor in 1958. [*Published with permission from American Philosophical Society.*]

A German refugee to Bristol in 1933 was the young E. Klaus Fuchs. Fuchs had openly been a member of the communist party in Germany during his student days and was the son of a Lutheran minister Emil Fuchs, who had Quaker ties.[63,64] In Bristol, Fuchs was supported by Ronald and Jessie Gunn and lived with them for a while, before moving to lodgings closer to the University. Ronald Gunn, who like Mott had attended Clifton College in Bristol, was a Quaker, a director of Imperial Tobacco, and Through marriage to Jessie, connected to the Wills family. Ronald Gunn was a suspected communist sympathizer with an extensive MI5 file.[65,66] The Gunns introduced Fuchs to Mott who accepted him as a Ph.D. student. It seems likely that the Gunns knew the Gurneys through their common interest in leftwing causes and Fuchs also became acquainted with the Gurneys. The Gurneys' association with Fuchs continued until his graduation from Bristol and move to Edinburgh in 1937.

Gurney's first publication in Bristol was his book *Elementary Quantum Mechanics* (1st Ed. 1934, 2nd Ed. 1940) (17) which he must have worked on in Manchester. In this book, Gurney illustrates the workings of quantum mechanics through a series of intuitive and semiquantitative arguments. He uses asymptotic and continuity conditions to show the form wave functions must take for systems with different potential energy functions; a method he called "fitting the psi-curves into the box", see Fig. 7. After intuition of the form of the wavefunction is developed, a mathematical solution to the Schrödinger equation is given. This book received a favorable review by the physicist Frank Hoyt.[67]

Gurney's books have a timeless quality and provide unique physical insights on each topic and this book is still useful as a supplementary source of intuitive understanding for a modern quantum mechanics or quantum chemistry course. In this respect, while being more specialized, they may be compared to the Feynman Lectures on Physics.

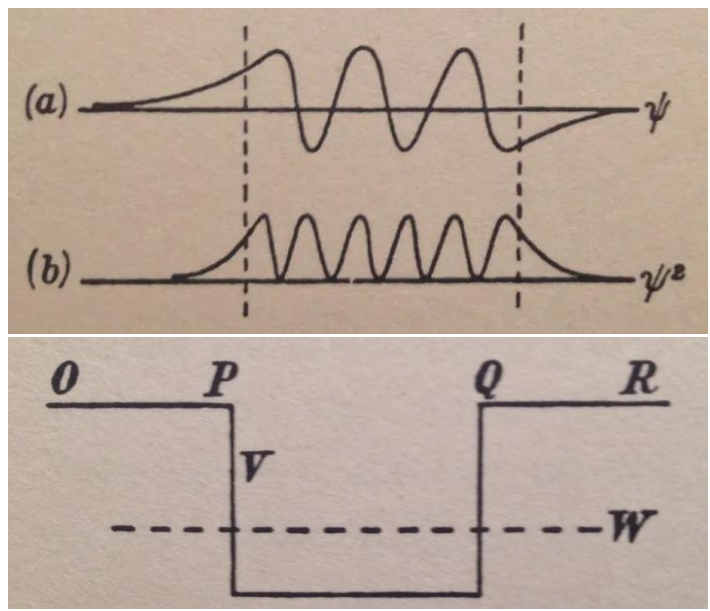

**Figure 7.** The square well potential and the qualitative description of the wave function (a) and probability (b) for the particle in the well. From Ref. (17).

At the suggestion of the Physics Department of the University of Bristol, Gurney sent his book to Leningrad State University for translation and publication. As payment, Gurney was offered Russian currency and as this could not be converted, in 1935 he and Natalie used the money to travel to Russia as tourists, visiting the Caucasus mountains region,[54] which at the time was a popular destination for foreign physicist visitors, including Dirac and Peierls.[30]

Gurney published his first paper in Bristol on "Theory of electric double layers" in March 1935.(18) This paper discusses the formation of an electric double layer on a metal surface upon adsorption of atoms of a second metal species. Gurney uses quantum mechanical arguments to show that independent of the magnitude of the work function of the adsorbed metal atoms compared to the electrode base metal, the adsorbed atoms can become positively charged at the surface.

Gurney's second book, *Ions in Solution* (1936) (19) expanded his recent work and gives clear descriptions of solution and electrode processes based on newly developed quantum mechanical ideas. This book was a primer for modern electrochemistry and illustrated Gurney's strong command of physical chemistry and experimental aspects of electrochemical cells. This book received a glowing review by the chemical thermodynamicist Frank H. MacDougall of the University of Minnesota who stated, "I have seldom read a scientific book with greater pleasure and interest than I experienced in going through this small volume by Dr. Gurney. … The treatment adapted by Dr. Gurney employs only the simplest mathematics but by frequent use of diagrams (particularly potential-energy diagrams) the author is able to give the reader an illuminating account of the phenomena under consideration".[68] This book was later updated and its scope expanded by Gurney's later book, *Ionic Processes in Solution* (1953).

Gurney's first published paper with Mott (1937) was an important contribution to solid-state physics of ionic crystals. In this work, they determined that the formation of *F*- or color-centers in ionic crystals are due to the capture of an electron by a vacant anion site in the lattice. (20) The nearest neighbor cations adjacent to the electron-carrying vacant anion site and next-nearest layer of anions relax to optimize their electrostatic interactions with the electron and form an attractive potential well, with bound quantized states for the electron. In a later paper, Gurney and Mott calculate the work required to form an *F*-center during the heating of alkali halide crystal in alkali metal vapor. (23)

Gurney and Mott published a landmark photochemical paper in 1938 where they give a modern quantum mechanical explanation of the century-old problem of the light-induced formation of latent images by silver salts on photographic plates.(22) They describe of how photoelectrons enter the conductance band of the $Ag^+$ ions in the AgBr salt and are then captured by secondary sites, perhaps consisting of specks of $Ag_2S$ on the surface of the AgBr grain which have lower-lying electronic energy levels. Interstitial $Ag^+$ ions migrate towards the negatively charged specks and are reduced to metallic Ag. Subsequent photoelectrons are energetically favored to reside on these Ag metallic sites, see Fig. 8(a), with the negative charge attracting further interstitial $Ag^+$ ions, Fig. 8(b). This leads to the formation of a submicroscopic grain of metal which seeds the latent image on the photographic plate.

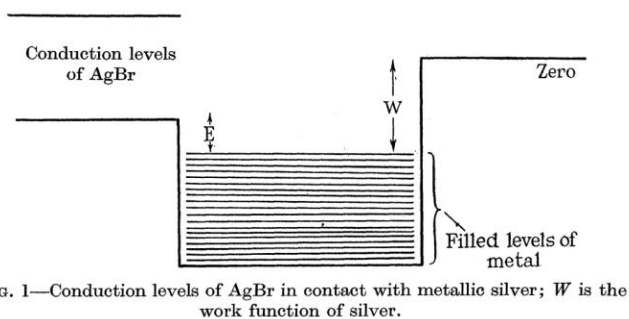
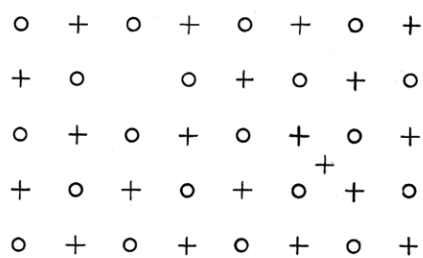

Fig. 1—Conduction levels of AgBr in contact with metallic silver; $W$ is the work function of silver.

Fig. 2—Displaced ion in AgBr lattice.

**Figure 8.** (left) The representation of the conduction levels of the AgBr ionic solid and the silver metal atoms formed upon photoreduction. (right) Interstitial, $Ag^+$ ions migrate towards centers of negative charge on the AgBr crystal. Ref. (22), with permission.

In 1937, Gurney published a review article on the theory of electrolyte solutions.(21) After qualitatively describing electrostatic interactions of ions in an electrolyte solution, Gurney summarized the statistical mechanical Debye-Hückel theory, and its extensions to higher ion concentration solutions by Bjerrum and Fuoss. He describes the electrophoretic effect of the ionic atmosphere on the motion of ions in weak and strong electric fields. The effects of salt concentrations on the viscosity, surface tension, volume, and dielectric constant of electrolyte solutions are covered. This is a modern review of electrolyte solution physical chemistry which can still be useful as a primer for students entering the field. Gurney published another paper on electrolyte solutions in 1938 where he used statistical mechanics to study temperature effects on electrostatic and van der Waals forces and how these affect dissociation reactions of weak acids and bases, including zwitterions, in aqueous media. (25)

With his increasing body of research in applied physics, for a few months starting in April 1937, Gurney travelled to the east coast of the United States to visit a number of industrial laboratories including General Electric, Kodak, and Bell Telephone to discuss and gather information on problems of mutual interest, maybe with the eye on future employment opportunities.[13]

In 1938, Gurney participated in a *Faraday Discussion* on reaction kinetics.[69] (23) He commented on why the heat of the recombination reaction of acetic acid in aqueous solution may be close to zero. The solution first heats up as the positive and negative ions approach each other, colliding with the solvent molecules along the way. After recombining, the electrostatic field of the ions disappears, allowing the solvent molecules to relax, causing a cooling of the solution. A balance of these effect determines the enthalpy of the recombination process.

In 1938 – 1939 Mott and Gurney surveyed the progress in liquid state theory in two papers. In the first paper Mott and Gurney summarized the statistical mechanical cell (hole) and distribution function theories of the liquid state which were active research topics at the time. They also summarized the quantum mechanical theories of the structure of liquid helium.(26) In the second paper Mott and Gurney criticize the use of the functional form of the configuration entropy in the Lennard-Jones and Devonshire hole theory of liquids and suggest an alternative form.(28) The hole theories may have been of interest to Mott and Gurney in the context of their studies of defects in solids.

In the theme of the photochemistry of solids, Gurney and Mott published another work in 1938 in a Faraday Society meeting which discusses factors effecting phosphorescence in photoexcited solids. (27)[70]

A lasting legacy of Gurney's time in Bristol is the book *Electronic Processes in Ionic Crystals* by Mott and Gurney published in 1940 (2nd Ed. In 1949). (29) In this book, Mott and Gurney include detailed discussions on dielectric properties of solids, types of imperfections in solids, the nature of electronic bands and holes in polar crystals, the theory of *F*-centers and formation of photographic latent images, and the quantum mechanical theory of semiconductors. They predict a formula for controlled emission currents flowing through dielectrics, including semiconductors, which is now known as the Mott-Gurney law. Reviews of this work were published soon after by Linus Pauling whose view was generally positive tempered with minor criticisms[71] and Philip A. Leighton whose view was positive.[72] Fröhlich later reviewed the second edition of the book.[73] This book became a standard reference for a generation of physicists and physical chemists.[see Ref. [74]: p. 66; Yoffe p. 166; Pepper p. 211; Thomas p. 257; Kamimura p. 291]

The mid and late 1930s saw the great depression, the rising power of the Soviet Union and Nazi Germany, the Spanish Civil War, and other geopolitical events which led many of the younger members of the physics department in Bristol and across the United Kingdom to become left-wing or at least curious towards the Soviet Union.[53,30] In Bristol, Mott, Ronald and Natalie, Cecil Powell, Stephen H. Piper and other members of the department and their spouses became members the benignly named Society for Cultural Relations with the Union of Soviet Socialist Republics (SCR USSR).[75,76,65] Natalie was honorary secretary of the Bristol branch of the SCR, coordinating the attendance of speakers with the London branch. In this role she wrote to a local newspaper in 1934 about the SCR's activities.[77,78] At the time, many prominent British artists and intellectuals including E. M. Forster, J. Huxley, H. G. Wells, Maynard Keynes, Virginia Woolf and Bertrand Russell were members of the SCR. Historians consider, VOKS, the Soviet home organization of the SCR, as a roof for Soviet Intelligence operations used for contacting foreign intellectual, scientific, and government circles. However, at the time, this may have not been clear to members of the society.[79] Mott and Fuchs attended one of the sessions of the Bristol branch of the SCR from 1936, vividly described by Mott: "At that time in the USSR Stalin was liquidating the men who had made the revolution, and verbatim scripts of the 'treason trials' were published in translation. … we read these trials there, each of us taking a part, and wondering innocently if these men really were guilty of the treachery imputed to them. Klaus Fuchs took the part of the prosecutor, Vishinsky, for which he threw away his shyness and attacked the accused with a venom which showed clearly enough where his sympathies lay." [53,80] Ronald and Natalie learned that members of the Soviet embassy attended some of the meetings and become disillusioned with the SCR and from 1936 – 1937 were no longer members. In later security screening and FBI interviews, membership in the SCR became a major problem for Ronald and Natalie.

By all appearances, Gurney was a valued and successful member of the Physics Department in Bristol. His work was central to Mott's research on ionic crystals and semiconductors. However, in 1939, the G. Wills Research Associateship was discontinued[51] and a permanent position did not materialize. Ronald and Natalie may have moved to London as early as 1938 as the Electoral Register for 1938 gives a London address for them.[81] Gurney had to move once again.

**The War Years and After (1939-1949)**

Gurney made another unconventional career choice when in 1939 he and Natalie moved to the Wenner-Gren Institute for Experimental Biology in the University of Stockholm as a visiting research fellow to study biophysics.[1,13] This institute had just been founded from donations by the Swedish industrialist Axel Wenner-Gren and the Rockefeller foundation and it was the vision of its first director, the Zoologist and Cell Biologist, John Runnström to bring quantitative experimental work to the study of biological function.[82] At this time, Mott had also arranged for Gurney to travel to various laboratories in Britain and abroad,[83] perhaps in the hope of finding a permanent position, but this plan had to be abandoned with the outbreak of World War II. Gurney was deeply depressed at the thought of another world war and was reluctant to return to England. The Gurneys initially intended to wait out the war in neutral Sweden. Despite mining of the waters of the North Sea in late 1939 during the period of the Phoney War, Natalie sailed home to England to visit relatives and friends for Christmas.[83,84] After returning to Stockholm, Natalie found employment as a secretary in the British Legation to assist in the war effort in whatever way possible.

Gurney immersed himself in biophysical research which resulted in two articles published in Swedish journals in 1940 and 1941. In the first paper (communicated in Oct. 1940), Gurney studied the rates of chemical oxidation chain reactions. (30) The paper showed that the reaction rate of the entire chain is equal to that of the slowest link, if the rates of the adjacent steps in the chain to the slow link are sufficiently rapid.[85] The second paper, cowritten with Runnström and another senior professor Erik Sperber, is a quantitative experimental study of the effect of the fluoride ion on the respiration and glucose metabolism of bakers' yeast. (31)

In June of 1940, with the fall of France, Denmark, Norway, and evacuation of British forces from Dunkirk, the Foreign Office directed British nationals to leave Sweden and if possible, return to the United Kingdom. As western sea routes were cut off and war on the eastern front had yet to begin, the Gurneys decided on an eastward route. After a frustrating six months of waiting for visas, in November 1940 Ronald and Natalie travelled from Stockholm to Leningrad.[86] They then took a train to Siberia and made their way to Japan. Finally, they sailed from Yokohama to the United States and arrived in Seattle on January 5, 1941.[1,83,87] After entry, the Gurneys travelled to California and stayed there for nearly two months, with Ronald lecturing at universities during this time.[86] Their first dinner gathering upon arriving in the United States was with J. Robert Oppenheimer in Berkeley, which rekindled a friendship which began in Cambridge. [88]

The Gurneys continued their eastward travel and arrived in Pittsburgh in February 1941. They stayed with Edward Condon and his wife Emily for about ten days while Ronald recovered from the flu.[2] At the time, Condon was the associate director of research at the Westinghouse Electric Corporation and working on the microwave radar program.[84] The Condons met Natalie [89] who Edward Condon described as vivacious and not at all shy. The Edward and Emily remained close friends and confidants to the Gurneys, perhaps strengthened by shared Quaker views. After Pittsburgh, the Gurneys went to New York City and Ronald reached out to faculty at Columbia University, including Joseph E. Mayer, Maria Goeppert Mayer [86] and the solution physical chemist, Victor LaMer seeking research opportunities.[90]

Gurney corresponded with R. H. Fowler and Charles G. Darwin, heads of the British Scientific Liaison in Washington, DC and in May 1941 the couple travelled there for direct meetings.[90,2] After contacting London, it was decided that instead of returning to England, Gurney conduct research in the United States. In late May 1941, Natalie secured an administrative secretary job, working with Dr. William L. Webster (a 1926 Cambridge Ph.D. and student of Rutherford), at the Washington-based Scientific Research Office of the British Supply Council in North America which dealt with wartime acquisitions. In Washington, Natalie became acquainted with the physicists Edward Teller and Ralph E. Gibson and their families. [91]

In the summer of 1941, Linus Pauling sent a letter to the American Commonwealth Fund strongly recommending Gurney for a fellowship[92] and Gurney become a senior research fellow from June 1941 to June 1943. [13,54]

In February 1942, the Gurneys rented an apartment near Columbia University in Manhattan and Joseph Mayer arranged for Ronald to be given an office there.[93] For part of his fellowship, Ronald worked at the Carnegie Institute in Washington, D.C. Natalie moved to New York,[94] but continued working for the British Science Office until 1943. [54]

For his fellowship, Ronald carried out experimental research for his future book *Ionic Processes in Solution*.(44) Gurney spent the summer of 1942 in Cold Spring Harbor in the Biological Lab, working on the book manuscript.[95] He stated that the goal of the book was to clarify the details of ionic process  in solutions, which could lead to better understanding of cellular and nerve physiology, and other biological and chemical systems. [95]

While preparing early chapters of this book, Gurney realized there was no "simple" and complete book on statistical mechanics of solutions to which he could refer.(44) Statistical mechanics textbooks available at the time were written for theoretically oriented physicists as intended readership. He put the writing of the first book aside and wrote *Introduction to Statistical Mechanics* with his typical intuitive approach. (39) We say more on these books below.

The Gurneys established personal friendships with many of the scientists at Columbia.[96] In New York Natalie and Ronald established an active and rich social life hosting a weekly "open house" with many colleagues from their professional circles.[1,2,21] True to her left-leaning politics, in New York Natalie became involved with social causes and joined the Union for Democratic Action (UDA) and hosted organizational events at their home. [94] The UDA was a pro-trade union organization whose additional goal was to advocate for the United States to assist other nations in their fight against Nazi Germany. The UDA was left-leaning but banned both conservatives and communists from its membership.

With the end of his Commonwealth Fund fellowship in 1943, Gurney was recruited to join the Ballistic Research Laboratory (BRL) at Aberdeen Proving Ground in Maryland by Oswald Veblen, with perhaps some facilitation by Joseph Mayer, who was a Consultant and Member of the Advisory Committee.[97] Gurney was hired at the BRL on a part-time basis in June 1943, but spent much of his time on the work there.

At the BRL, Gurney applied classical mechanics to ballistics problems, specifically to the fragmentation of explosive projectiles, and published a monograph on this topic rather quickly in September 1943 (32). With his typical physically intuitive approach, Gurney used energy balance between the chemical energy density of the explosive and the kinetic energy of the metal fragments of the shell to determine the maximum speed of shell fragments. The model was later extended to include momentum balance and is considered a classic in the field of ballistics. His analysis of the process is called the Gurney model and Gurney parameters are still used to characterize the explosive output of driven metal shells.[98-101] Gurney was very productive at the BRL, publishing six widely-used and cited reports.(32-37) [13] Many prominent scientists worked at the BRL at the time including S. Chandrasekhar, J. von Neumann, M. Schwarzschild, L. H. Thomas, R. G. Sachs, and E. Hubble and Gurney would have interacted with them.[102] There were more junior staff including physicist William Sleator Jr., and Richard N. Thomas who later became prominent scientists. In some cases, staff members became family friends who attended Natalie's salons.

Ronald later composed a cello and viola piece for Sleator and his wife Esther who was a medical doctor.[103] The dedication of this piece, written in Restoration Style in Ronald's handwriting is shown in Fig. 9 and gives a glimpse of Ronald's more private personality. Ronald also composed a piece for four instruments which Esther and William shared in performance. [104]

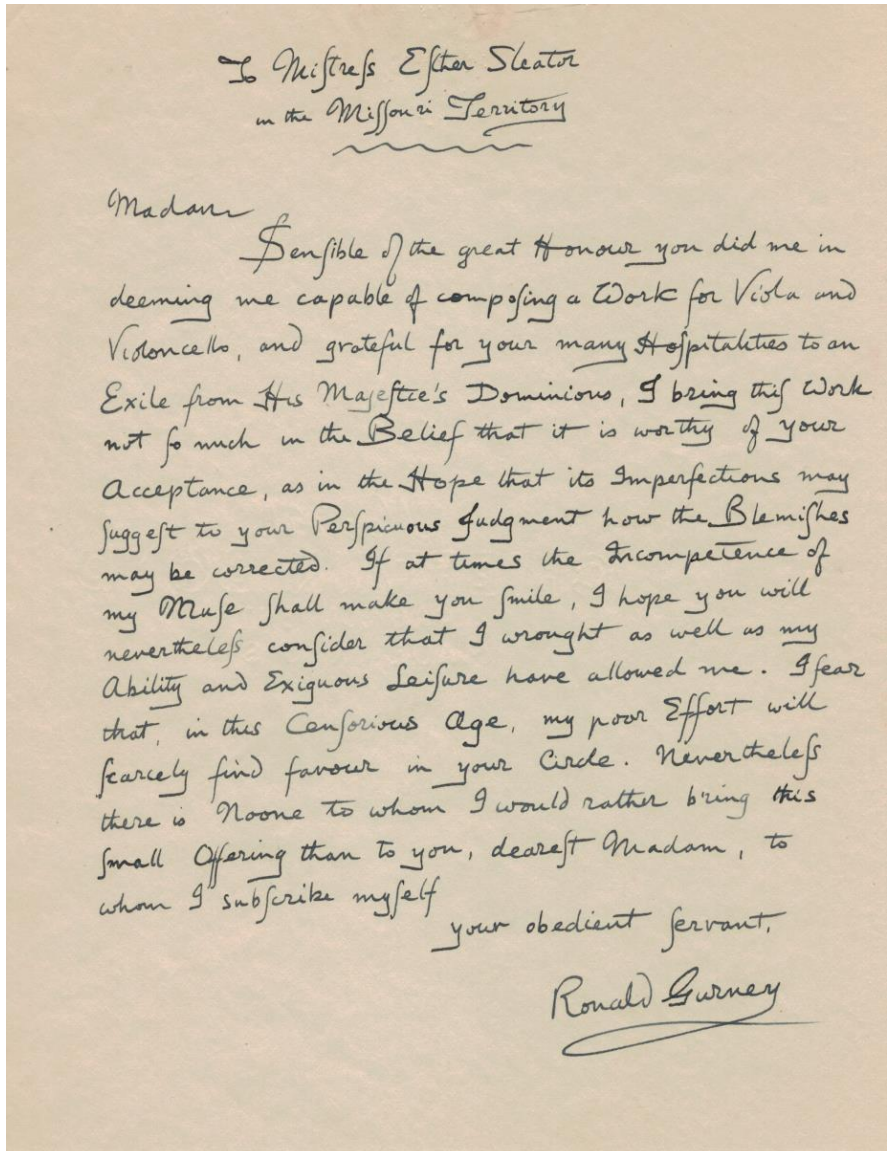

**Figure 9.** Ronald Gurney's handwritten dedication of his cello – viola duet to Ester Sleator. This letter and the music score were mailed on Mar. 21, 1950, under difficult circumstances for the Gurneys, see below. (Courtesy of Prof. D. Sleator).

Based on her interest in international affairs, Natalie became employed in the Southeast Asia Institute in New York City some time in 1943. The president of the Southeast Asia Institute t was the Dutch – American scholar Prof. Adriaan J. Barnouw, the Queen Wilhemina Lecturer at Columbia University,[105] whom Natalie later describes as her "second father". Natalie became a member of the Board of Directors (which included Margaret Mead) in the role of Executive Secretary.[106,107] From 1943 to 1947, Natalie also joined the Institute of Pacific Relations.[54]

In 1943, Gurney stated that he did not intend to stay in the United States after the war and as early as 1944, he stated interest in returning to the University of Bristol after the war. However, lingering resentment over his failure to return to Bristol earlier and the perceived lack of noteworthy accomplishments since leaving Bristol made this impossible.[83,108] His work in Stockholm was likely not considered, the new books he was writing had yet to be published, and his work at the BRL was confidential and likely not of high profile, as work on the Manhattan project would have been. Perhaps with no position materializing in England, Gurney spent August and September 1945 in Canada to obtain an immigration visa to the United States, as the first step towards eventually applying for citizenship in April 1947.[54,77]

With the end of the war, Gurney travelled to England in September 1945 to work in the Ordnance Department of the United States Military Attache in London until March 1946.[56]

**Chicago (1946-1948)**

In the summer of 1945, Gurney was offered a job at the University of Chicago, Metallurgical Lab (later named as Argonne National Lab) as Chief Physicist in the Theoretical Nuclear Physics Division. This was a welcome opportunity for him to return to nuclear physics research. He was given the requisite additional security clearance by the FBI and Military Intelligence and in November 1946 the Gurneys moved to Chicago and found an apartment close to the University of Chicago.[109] Gurney conducted experimental work on the pile,[110] studying scattering of slow neutrons by polycrystalline solids as part of the Manhattan Project.(38) During this period, he was occasionally called in to the BRL to address special problems.[77] The Gurneys stayed in Chicago until 1948 and stated a distaste for the city compared to New York. [110,111] In Chicago, the Gurneys once again established an active social life and friends among their circle included the physicist James Franck, the Manhattan Project metallurgist Cyril S. Smith, the Tellers, and Mariette Kuper, the first wife of John von Neumann.

Sometime in the summer of 1947, the Gurneys vacationed in Mexico and re-entered the United States at Laredo Texas border on September 7.[20] Klaus Fuchs and his brother-in-law Robert Heineman, a later suspect in the Fuchs case, made trips to Mexico prior to this and the Gurney's later raised the suspicions with the FBI. Fuchs had travelled with, Rudolf and Genia Peierls, and Augusta Teller to Mexico City in December 1945. This was a parting trip before the British party at Los Alamos returned home.[112,113,80] Heineman travelled to Mexico between February to August 1947 and was documented to have attended university Spanish language classes and visited a branch of the National Archives in Monterrey, which aroused suspicion. Heineman also vacationed during this time.[114,115] At the time, travel to Mexico was of concern given that Soviet agents were known to be active there.

In November 1947, the Gurneys who lived across the street from Edward and Augusta Teller, met the visiting Klaus Fuchs at the Teller home. Augusta Teller invited friends to their home to meet

Fuchs.[116,117] By then, Fuchs, aged 36, was Head of the Theoretical Physics Division of the Atomic Energy Research Establishment (AERE) at Harwell, UK, and Gurney commented that achieving such a high position at a young age had made him conceited.[20] Their contact with Fuchs during his stay in Chicago time was limited to Fuchs talking with Natalie about mutual acquaintances in England.

In 1947, Brookhaven National Lab was established with Philip Morse, Ronald's coauthor from Princeton, as its first director. Morse offered Ronald a scientific and Natalie a secretarial position at Brookhaven.[118,119] They accepted the offers and the couple even threw a good-bye party for their Chicago friends and returned their apartment. However, after some back and forth, they could not get the requisite Department of Defence security clearance for Brookhaven, despite maintaining the Argonne security clearance.[1] With difficulty, they found a new apartment and Ronald resumed working at Argonne.[120] It was not clear to Edward Condon why the Brookhaven security clearance did not come through as his impression was that Gurney was totally apolitical and additionally, research at Brookhaven was not planned to be on classified topics.[21] Later documents indicate the denial of security clearance was due to concerns about Natalie's loyalty, mostly related to her association with the SCR and refugee support groups, and perhaps her current United States associates.[120,121,20] Natalie defended herself against charges of being a communist or being sympathetic to the Soviet Union and submitted supporting character letters from leading Quakers in England, Lord Marley former secretary of the Student Christian Movement, and Charles G. Darwin.[118]

Some time during his work at Argonne, Ronald underwent a medical examination and was informed he had a serious case of hypertension. He only made a casual reference of this examination to Natalie, but at the same time started making arrangements to leave Argonne and return to academic life.[13] The Gurneys became frustrated with the clearance process at Brookhaven and withdrew their application. In hindsight had they persevered, they would have likely been able to get clearance for Natalie at this stage. From their letters and general decision-making process, it appears that the Gurneys did not have a strong grasp of the workings of the government and university systems in the United States and had not cultivated relationships with influential high-ranking government or university officials. Ronald's image is that of a quiet researcher immersed in scientific work, while Natalie is fully engaged in social causes, with left-wing leanings. Most of Natalie's character letters were from left-leaning individuals who, given the atmosphere of the times, would likely not have been effective in clearing her name. Neither of them seems to have had an acquaintance who could give timely advice on security matters and often it was some time after trouble had precipitated (often beyond repair) that they would ask for advice and help through letters to their friends.

Ronald started negotiating with universities for faculty or research positions and was offered a two-year visiting professorship at Johns Hopkins University, which included teaching and research duties. Ronald had a permanent position at Argonne and the implication was that after the two-year contract, he would be made a permanent staff member at Johns Hopkins. The Gurneys left Chicago for New York in June 1948 and moved to Baltimore in September 1948 and Ronald started working at the Physics Department.

**Baltimore (1948 – 1952)**

At Johns Hopkins, Ronald taught quantum mechanics and other topics. Louis Witten (father of the physicist Edward Witten), a graduate student at Johns Hopkins at the time, had the impression that Gurney was a good physicist but not a good teacher![122] Given the clarity of his writing this is surprising and could have been due to his shyness, health situation, or preoccupation with their security situation.

Almost immediately upon arrival in Baltimore, in Oct. 1948 Natalie began studies towards a graduate degree in International Relations at the Walter Hines Page School of International Studies with Prof. Owen Lattimore. Two of Natalie's acquaintances from the Southeast Asia Institute, John K. Wright (Director of the American Geographical Society) and Ralph Linton (Yale University) and also Condon sent letters of recommendations to Lattimore on Natalie's behalf.[56,123] Later Lattimore came under investigation by the FBI for pro-Russian leanings and Natalie's association with him became a further subject of questioning for both Ronald and Natalie, particularly since they had attended a dinner meeting at the Lattimore home where discussion had centered around developments in China.[56] In Baltimore, the Gurneys expanded their social circle which included the physicist Gerhard H. Dieke and Richard T. Cox and their families. They bought a car and all seemed to be going well.

On February 16, 1949, possibly due to pent-up stress from the previous year and his extreme hypertension, Ronald felt faint near the end of a lecture and collapsed from a stroke a short time later.[122,124] He was rushed to a hospital and spent several days in intensive care. After five weeks of therapy and exercise in the hospital, he made a remarkable recovery.[124,12]

While recovering, Gurney's book *Introduction to Statistical Mechanics* (39) was published, which provided a welcome distraction. *Introduction to Statistical Mechanics* had a unique approach to the teaching of statistical mechanics. The usual route used in textbooks is to assume knowledge or at least appreciation of theorems in probability theory for distributions with large numbers of variables and other advanced mathematical physics techniques. These mathematical results are then adapted to derive the most probable distribution for the system / ensemble among available energy levels, subject to macroscopic constraints put on the physical system. Gurney's alternative is to introduce a tractable example of distinguishable, weakly interacting classical systems with equally spaced energy levels. He determined that for a fixed number of particles and total energy content, averaging over all possible distributions of the system, subject to constraints, gives a "graded" distribution for the occupancy of the energy levels, in which each level of lower energy has a higher occupancy than levels with energy above it. A simple calculation proves that for systems with large numbers of particles, the graded distribution with highest probability is the exponential decay (see Fig. 10), with the decay constant inversely proportional to average energy per particle in the distribution. The most probable distribution with exponential form is shown to be overwhelmingly more probable than all but a very limited number of other graded distributions. This explicit illustration establishes an intuitive framework for which more general approaches, which use ensembles and results from probability theory of large numbers can be derived. After discussing the basic concepts, the book covers a set of examples, some based on Gurney's own research on alloys and solutions. (39) Noting its intended limited scope, this book received a positive review from the physicist David M. Dennison of the University of Michigan [125] and the physical chemist Pierre van Rysselberghe of the University of Oregon.[126]

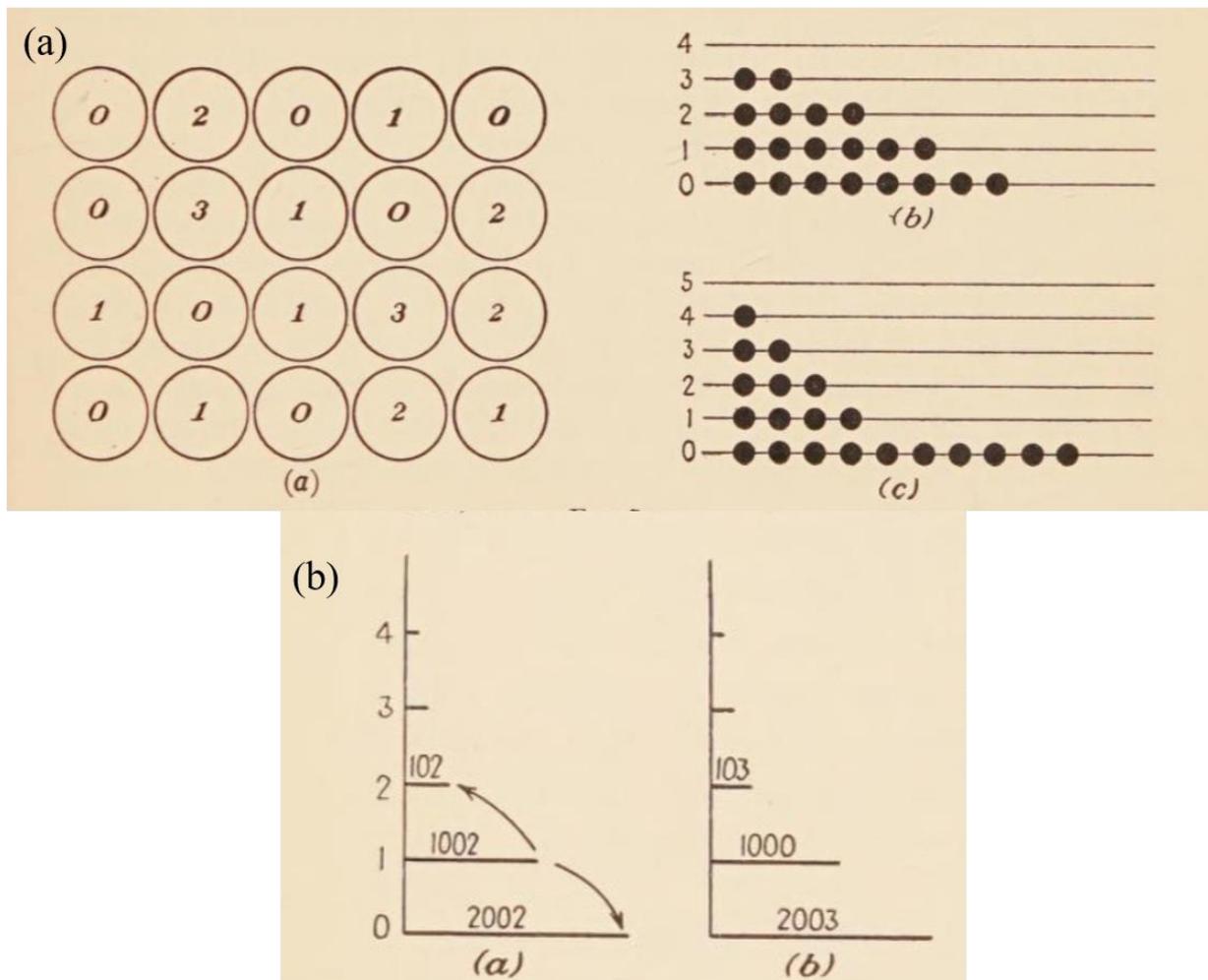

**Figure 10**. (a) Left, the occupied energy level (indicated by numbers) of 20 distinguishable molecules (shown by circles). Right, two high-probability graded distributions of the 20 molecules among energy levels. (b) The set-up used to prove that the most probable distribution is graded with an exponential decay form corresponding to exp(-$\beta E$), where $\beta$ is a decay constant related to the inverse of the average energy of the distribution. [From Ref. (39)]

Ronald convalesced with Natalie and friends in Maryland in April and May. In June 1949, the Gurneys travelled to England for a summer vacation to stay at Ronald's family home. While in England, at the invitation of the Director, Sir John Cockroft, Gurney visited AERE at Harwell, likely to seek a permanent position. Cockroft could not be present during Gurney's visit and asked Herbert Skinner, the Deputy Director of AERE to show Gurney around. Skinner delegated the job of showing Gurney around the establishment to Fuchs, who was the head of the Theory Division. Other than this encounter, the Gurney and Fuchs seem not to have spoken.[20,127] We find later that Gurney was denied a job at AERE based on unspecified "loyalty considerations."[20] The visit in September was unfortunate as Fuchs was then under increasing suspicion of spying, followed by his arrest in February 1950. This meeting raised intense suspicion on the Gurneys.

After England, the Gurneys spent two weeks in Monaco with Virginia and Richard Adloff who was an expert on the Far East. They returned to the Baltimore in early October 1949.

Unbeknownst to the Gurneys, during Ronald's illness and recovery, the next storms were brewing. A "Loyalty Investigation" for Ronald was instigated by the FBI in May 1949 at the request of the Department of the Army. This investigation was discontinued when his employment at Aberdeen Proving Ground was formally terminated in June 30, 1949. However, the Army, Navy, Air Force Personnel Review Board on Sept. 15, 1949, denied Gurney access to aeronautical or classified contract work or information.[20]

In parallel, subsequent to Gurney's illness, a Kafkaesque stream of "misoccurrences" ensued with regards to his employment at Johns Hopkins.[128] After Gurney's stroke, his doctor wrote a note advising that it would not be good for his health to lecture again and forwarded this note to the University. The physician had meant this as a general recommendation to avoid stressful situations, but Johns Hopkins University took this note literally and terminated Gurney's contract while he was on medical leave and hired a replacement in September. To their credit, the university arranged for Gurney to have a research job at the university-affiliated Applied Physics Laboratory at Silver Spring, Maryland, which did research for the Navy. However, this job required a security clearance which as described above, Gurney had been denied. After recovery and time away in Europe, Gurney returned to find out he no longer had a job, and furthermore, many of his colleagues were unaware of his predicament which had occurred while he was away. Rather than being vocal and trying to resolve the issue of their security clearance with the Navy, the Gurney's were quiet, not wanting to attract attention to the fact that their security clearance was denied. This was a big blow to Ronald and what made it more difficult was that he had left the permanent position at Argonne, with the understanding that the position at Johns Hopkins would become permanent after two years.

On Condon's recommendation, Gurney applied for a position as "Physicist – Sensitive" at the National Bureau of Standards (NBS) in Washington DC in Dec. 1949 and yet another FBI Loyalty Investigation was launched. While they were waiting for the outcome, Ronald's mother died just after Christmas 1949, and he left for England in January 1950 to be with his sister and settle family affairs.[129] He returned to the United States in March.

News came late in January 1950 that Ronald was again denied the security clearance for the NBS based on Natalie's perceived communist connections and/or sympathies. This time, the loyalty investigation included more serious allegations of association with Klaus Fuchs. To quote directly from the letter to Ronald, "Your wife has been consistently associated with organizations which are Communistic, or follow the Communist Party line, and she has consistently and sympathetically supported the activities of such organizations which have been detrimental to the best interests of the United States Government".[60] It seems that the material Natalie had submitted earlier in 1948 to clear her name for the Brookhaven position had not been considered. Ronald was given a chance to appeal in Feb. 1950 and to spare Ronald the stress, Natalie appeared on his behalf at the security hearing before his return from England. She obtained more than 30 letters from academic and cultural friends and acquaintances in the United States and the UK testifying to her loyalty.

When Gurney returned, he was informed that further "field work" was required before a decision could be made on the security clearance for the NBS. The Industrial Review Board initially upheld the decision to deny the security clearance on March 1950. With further follow-up with the review board by Natalie and letters of recommendation from Oppenheimer, Urey, and others,[130] the Industrial Employment Review Board granted Ronald a security clearance for the type of work he was doing before. The report for the updated loyalty investigation dated May 1950 involved numerous interviews with Ronald and Natalie (see next section).[20] The verdict quoted by Gurney in a letter to Oppenheimer was: "The denial to the appellant by the Army-Navy-Air Force Personnel Security Board of the privilege of access to the Department of Defense classified information and contract work was effected without sufficient cause, and the same is hereby reversed."

Despite this turmoil, in June 1950 Natalie earned an M.A. degree in International Relations with her essay on: "History of the Territorial Dispute between Siam and French Indo-China and Post-War Developments in the Disputed Territories."[131] Her supervisor Owen Lattimore was himself subject of a McCarthy era investigation and this later implicated Natalie by association. [20] Natalie originally intended to get a Ph.D., but under the circumstances, had to finish with an M.A. degree.

The Gurneys once again travelled to England between June and Sept 1951. During this period, they spent a few days with Ronald Gunn.[65] Later, in a letter to Klaus Fuchs in prison, Gunn wrote about the Gurneys', "Ronald and Natalie Gurney have recently stayed with us a few days. No doubt you remember them. Ronald had an illness a year or two ago but seems to have made a good recovery. Natalie is as vivacious as ever and could get away with anything short of murder. He has been writing some scientific books, alone or in collaboration. Natalie has got an M. A. degree." The fact that Gunn mentions the Gurneys in a letter to Fuchs would imply to an outsider that their acquaintance had been close. This encounter with Gunn and the letter, which was of course read by the Security Service, may have been the trigger for a new set of allegations against the Gurneys.

Due to the delays in gaining the National Bureau of Standards security clearance, Gurney sought other work and was hired at the Institute for Fluid Dynamics and Applied Mathematics at the University of Maryland some time in September 1951. At this junction Gurney's health was quite poor and naturally his scientific output suffered. In 1951, as part of the University of Maryland lecture series, he published a report "A Problem in the Statistical Mechanics of Alloys".(40)

In 1952, Gurney published three papers. The first paper, with coauthors from the Aberdeen Proving Ground, is an experimental / theoretical work on spectroscopic measurement of the equilibrium constant for acid – base proton transfer between bromocresol green and triethanolamine in non-aqueous solutions methanol, ethanol, and n-butanol. (41)

Gurney's final two of the papers describe his work at Argonne National Laboratory which was declassified and could now be published. (42),(43) In the first paper, to better understand the loss of energy by slow neutrons in graphite, Gurney studied the vibrational spectrum of the two-dimensional Debye model of the graphite lattice. (42) The second paper studies the problem of the cross section for energy transfer of slow neutrons to lattices described by the Einstein model and shows that the efficiency of moderation of slow neutrons improves for lattices made up of lighter atoms. (43) Computations for this work at Argonne were conducted by Marvin L. Goldberger, who was a Ph.D. student of Enrico Fermi,[132] and after a distinguished career as a theoretical particle physicist, became the president of the California Institute of Technology.

Another unfortunate and unnecessary incident triggered the last chapter of Gurney's story. Some time in 1952, with respect to his duties at the Institute of Fluid Dynamics, a colleague Raymond J. Seeger thought it useful for Gurney to get a security clearance from the Navy to visit the Naval Ordnance Lab.[133] Gurney did not need to visit the Navy labs, but as a matter of routine, Seeger thought this would be useful. However, despite having a recent security clearance based on Natalie's submissions, a security clearance to visit Navy sites was once again denied. Initially Gurney was not concerned since the University of Maryland had plans to open the Institute of Molecular Physics headed by the Dutch physicist Antonius M. J. F. Michels and they were anxious to get Gurney onboard as a permanent staff member for non-classified work. [133]

However, subsequent to the latest Navy investigation, an FBI agent visited the President of the University of Maryland, Harry Clifton Bryd, telling him of Gurney's (and that of other faculty at the Institute of Fluid Dynamics) denial of clearance. At Bryd's orders in April 1952 Gurney and five other professors were expelled from the University of Maryland without any hearings. Ronald and Natalie sent letters to the Navy and got his clearance reinstated, but the University President would not change his ruling and this was backed by the University Board of Regents. Ronald's position at the University of Maryland was terminated.

Ronald was in a bad state of mind and felt "hunted".[133] The Gurneys stayed in Baltimore until June 1952. Ronald decided to go to England from July to September 1952 and Natalie went to their apartment on Riverside Road in Manhattan, which they had been subletting all this time.[134]

### The Klaus Fuchs affair and interviews with the FBI
As early as 1937, Natalie had been questioned by Scotland Yard in Britian in relation to her work with refugees from Nazi Germany. By 1940, Ronald and Natalie had been assigned a MI5 personal file (PF) number as potential persons of interest (PF 42,397), however, there are currently no extant files on the Gurneys in the UK National Archives. Their interactions with Ronald Gunn were monitored. These concerns where not deemed significant enough and they were both given security clearances to work with the British Embassy in Washington, in Army facilities at the Aberdeen Proving Ground, and at Argonne National Lab.

After his arrest in 1950, Klaus Fuchs was interviewed extensively by MI5 and the FBI and there were mentions of the Gurneys. In FBI documents, it was established that Fuchs was associated with the Gurneys in Bristol until 1937.

In FBI interviews Fuchs stated that he regarded Ronald Gurney as a security risk because he had been denied the position at Harwell due to loyalty concerns. More so, he considered Natalie a security risk since she was indiscreet and based on her extensive social contacts, was vulnerable to inadvertently becoming an intermediary between Soviet agents and sympathetic individuals in the west. Fuchs said that he had no knowledge of the Gurneys undertaking espionage activities for the Soviet Union, and his conclusions were based on the Gurneys being members of the SCR which he considered a Soviet spying front.[19,20,134] Fuchs' offhand remarks are repeated in FBI documents from 1950, 1951, and 1952 and in non-redacted segments of the documents, are treated as damning evidence, with no further corroborating evidence offered.

The Gurneys were also extensively interviewed by the FBI in relation to the Fuchs case. Natalie was the main subject of interest in these interviews and FBI interviews with Ronald are substantially briefer. The FBI asked Ronald about his membership in the SCR in Bristol and he responded that he liked to learn about other countries and had already made a number of international trips prior to his time in Bristol. There was no mention of Ronald ever being a member of the Great Britain Communist Party or his trips to the Soviet Union in 1932 and 1935. Ronald states that he left the SCR in 1936 and Natalie left it in 1937 as they became disillusioned with the goals of the society.

In her interviews, Natalie was accused of communist associations. Natalie admitted that in 1937, she was interviewed by two Scotland Yard men who informed her that a communist refugee, who attempted to land illegally in England, had a letter of introduction to her among his papers. She stated that she had never learned the identity of this individual and had no prior knowledge of this. Natalie said that some German refugees who came to her organization for assistance were sent with letters of introduction from the Quaker organization, but that she did not recall any instances where a person had secured such a letter prior to leaving Germany.

Natalie advised that she was not aware that any of her social acquaintances in New York City belonged to organizations that were considered pro-Communist and that she had never been invited to attend any meetings of organizations which were pro-Communist by persons who visited her home. Natalie mentioned that it was customary for her to open house once a week while she resided in New York City, but guests were, for the most part of scientists from the Southeast Asia Institute, the Commonwealth Foundation, and Columbia University.

Natalie recalled Klaus Fuchs as a student at Bristol University in 1935 and 1936 and that he was on the fringe of her social set. She recalled several instances where he was part of a social group which included herself and Ronald but was unable to recall specific information concerning the number or nature of these gatherings. She advised that during the time she knew Fuchs in Bristol, she had never observed any conduct indicating he was a communist.

In FBI files, there is a list of 18 names (with further names redacted) who were interviewed with respect to the Gurneys.[20] The FBI went as far as surreptitiously obtaining handwriting and typewriter samples from the Gurneys and obtained a photo of Natalie from the Registrar's office at Johns Hopkins where she was studying. Their house in College Park was under surveillance from January to April 1951. [20]

At the end of 1951, these FBI documents do not find evidence of "communist activity or pro-Russian or Soviet sympathies." There were suspicions based on hearsay and the Gurneys' trip to Mexico seems to have aroused suspicions. The final conclusion was: "It is believed, in view of the results of investigation to date, that these two non-specific, unsupported allegations or opinions are not sufficient basis for a full-scale espionage investigation of Natalie Gurney at this time." [20]

**A tragic ending**

In New York, their troubles did not end. Natalie wrote that from September 1952, there were yet another set of allegations against Ronald, this time, distinct from the case against Natalie.[135] These may have arisen from Ronald's visit to Harwell which was construed as "an unexplained and unnecessary visit to Klaus Fuchs", even though the visit was at the invitation of Sir John Cockcroft.[136] The allegations may also have arisen in relation to his time at Argonne, or from his name appearing in an open FBI investigation on Harold Urey.[137]

In New York City, whether out of fear or lingering suspicion, his friends and colleagues in the Columbia Physics and Chemistry Departments seemed to not welcome his return, to the extent of denying him library access, not inviting him to seminars, and not sitting with him during lunch. This impacted Ronald deeply. Condon upon hearing of this from Natalie wrote a harsh letter to the Columbia faculty and his Physics library access was renewed. However, even after gaining library access, he was still socially isolated by colleagues. They did not interact with him and despite prior family contact, declined invitations to visit the Gurneys at their apartment. In December, without Ronald's knowledge, Natalie wrote letters to I. I. Rabi and L. P. Hammett asking why they were treating her ill husband in this manner despite previous bonds of friendship.[138]

In the absence of academic employment, Gurney secured consulting work with the Corning Glass Company where Condon was the Research Director.[139] At this time Condon was himself having troubles with security clearances. Gurney also did consulting work with the Milwaukee Gas Specialty Company, and the Freed Radio Corporation. His colleagues at the BRL at Aberdeen were hopeful of getting him to collaborate with them again.[13] Despite everything, Gurney's scientific contributions were obviously still valued. Due to his health Natalie assisted him in completing his book, *Ionic Processes in Solution*.[3]

In spring 1953, Gurney was invited by the director of the Kodak Research Laboratories to visit their facility in Rochester, New York, and work with them. He was also introduced to staff at IBM with the view of a possible consultant job in the solid-state field. He gave a lecture to IBM staff on March 30, 1953 and was excited about lecturing after three years and seemed fine afterwards.[13] A few days later, he had a retinal haemorrhage and was told he should be hospitalized for tests. However, he only stayed home and rested his eyes. He talked to Natalie about writing a sixth book on "the Theory of Electronics" and made notes on topics to include in this text.

Gurney had a stroke in the middle of the night in their apartment in New York City on 14 April 1953. A doctor arrived quickly but he died within a half an hour of collapsing. The cause of death was announced as a cerebral haemorrhage. [13,133]

Natalie was very distressed by Ronald's death. She learned that security agents contacted the doctor who arrived at the scene of Ronald's collapse to determine if he had spoken any words at the time of death. Thus in addition to the blow of Ronald's death, Natalie assumed there was an active case against her and she was under surveillance and possibly subject to arrest. She was at loss and wrote to Condon, Seeger, and Morse for assistance but did not receive any replies. Natalie tried very hard to clear their names with security officials. The mathematician James R. Newman introduced Natalie to a lawyer who she consulted regarding the security situation with the FBI.[140]

In the days following Ronald's death, Natalie received as many as 100 letters of condolence and praise of Ronald's person and contributions from scientists whom she personally did not know well. Letters of sympathy arrived from Col. Alden P. Taber, Director of the BRL at Aberdeen Proving Grounds on behalf of himself and the staff, [141] Prof. Melba Phillips of Brooklyn College, Dr. Kenneth Mees of Eastman Kodak, Prof. J. C. M. Brentano of Northwestern University, Prof. Franco Rasetti of Johns Hopkins, Prof. James Franck of Chicago, and Prof. A. Barnouw, Natalie's "second father" from Columbia.[34] Natalie communicated extracts from some of the letters to other friends and these are reproduced in the Appendix.[142,143] To her considerable anger, Natalie did not receive a single letter from Ronald's friends and colleagues at Columbia, who she called, "the league of the fearful".[144] Close friends, Morse, the Tellers, and the Mayers had not written, but what Natalie did not know is that some of these friends were subject to security investigations of their own related to the Fuchs affair.[145] Natalie again wrote strongly worded letters to Rabi and Hammett protesting the way they shunned her ill husband and how it contributed to his death.[138] Despite maintaining contact with Emily Condon, even their close friend, Edward Condon did not acknowledge Ronald's death to Natalie. It took a strongly worded letter from Natalie asking the reason for his silence,[146] for Condon to write in July 1953. Condon expressed condolences and wrote about his own serious problems with security clearances.[144] He later regretted not writing Natalie and offering support sooner.[2]

Obituaries for Ronald appeared in the *New York Times*[147] and the *New York Herald Tribune*,[148] both which contained a surprising amount of detail about Gurney's contributions and were written by someone with technical information on Gurney's work. Natalie sent letters to Sir Lawrence Bragg, Sir Nevill Mott, Sir John Cockcroft, Edward Condon, Eugene Rabinovitch (from the Bulletin of Atomic Scientist) and journals *Physics Today*, *Science*, and *Nature* providing biographical information for them to publish an obituary or short biography on Ronald. In these letters, she included Ronald's obituary from the *New York Times* and *New York Herald Tribune*, Ronald's biographical outline which she had prepared, extracts from letters of sympathy received from Ronald's associates,[34] and an outline of Ronald's recently completed book on ionic solutions.[13]

The only signed obituary for Gurney published near his time of death was in *Nature* by Sir Nevill Mott.[3] By a curious coincidence, in this issue of Nature, Gurney's obituary is followed by that of the philosopher C. E. M. Joad, in whose Progressive League Natalie and Ronald met for the first time.[149] A short, unsigned obituary appeared in *Physics Today*,[150] and his death was noted in *Science*.[151] There was no memorial service held in Columbia. Natalie thought that one day a suitable memorial would be in the form of a publication or a fellowship at Cambridge or Princeton in his name.[136]

Gurney's final book *Ionic Processes in Solution* was published posthumously in 1953.(44) This book was written from a physicist perspective for electrochemists and researchers in the field of ions in gas and liquid phases and the approach is different from physical chemistry texts on this topic. The first six chapters summarize electrostatics, thermodynamics, transport properties, and statistical mechanics of ionic solutions and include discussions of solutions of ionic salts, solutions adjacent to electrodes, and solutions of weak electrolytes. The book continues with a summary of experimental results, some of which are from Gurney's own work between 1941 to 1943. There are discussions of proton transfer, solvent effects, and dissociation of electrolytes at different concentrations. This book received a good review by Sydney M. Selis of the University of Maryland[152] who stated, "Ronald W. Gurney is an excellent teacher. His novel approaches will be appreciated by the student as will be his extremely lucid descriptions. When introducing new subject matter, he carefully reviews what has come before, and in doing so, he anticipates the difficulties which the individual student will meet."

As things settled, Natalie realized it was likely pointless to apply for jobs in the United States in her field of International Relations, for which she would need security clearances.[153] She returned to England in May 1954 and stayed with her sister-in-law at the Gurney family home in Cheltenham. She bought a house in Hampstead, London in May 1955.[141] Natalie continued corresponding with the Condons, Oppenheimer, James Franck, and other friends from the United States after Ronald's death. The Condon's son Paul visited Natalie in her home during the period he was studying with Sir John Cockcroft in London, and Edward and Emily Condon visited Natalie during a trip to England. The last correspondence on record between Natalie and Edward Condon is from 1973, a year before his death.

In 1958 Condon and Hugh Odishaw published *The Handbook of Physics* which includes a chapter entitled Ionic Crystals written by Gurney.(45) Condon had this project in mind for a long time and Gurney gave him a draft of the article after they moved to New York in 1952. A point of disappointment with Natalie was that Gurney's affiliation in this article was given as the University of Bristol. While Condon tried to explain this as an editing error, Natalie stated that after all these years, it seemed that there still was fear of being affiliated with Ronald in American universities.[155]

In 1958, Natalie married the Czech-Austrian psychiatrist Fredrick Kräupl-Taylor (Fritz Kräupl, 1906 - 1989) (see Fig. 5) and changed her name to Natalie Gurney-Taylor. [156,157] She was active in maintaining Ronald Gurney's scientific legacy and prompted by her family friend Paul Rosbaud,[158] (the co-founder, with Robert Maxwell, of Pergamon Press) and  British spy in Nazi Germany,[159] she renewed the copyright of two of his books and got them reprinted.[158] She kept in touch with the physics community in England and with an introduction by Condon, she and her husband met and befriended David Bohm and his wife Sara in 1961.[158] She hinted at writing a private book about her and Ronald's experiences, but there is no further record of this. Natalie lived a full life in England, which merits a treatment beyond the scope of this article. It has not been possible to locate any of Natalie's relatives and it is not known if archives or collections of her letters or photos of Ronald Gurney are still in existence.

Edward Condon, for different reasons did not write an obituary or tribute article to Gurney at his time of death, but paid tribute to his friend in two impactful articles, the first published in 1973 [1] and the second published posthumously in 1978.[2] These inspired a further article about Gurney's contributions to ballistics calculations.[99] Condon also discussed Gurney at length in his AIP Oral Interview[21] and described him as "A very, very brilliant guy, but not brilliant in the Feynman sense. He was as brilliant on paper and as careful a guy about papers and understanding. ... But, you know, just too shy to hold a teaching job or be a lecturer. I don't suppose he ever directed a student to get his doctor's degree and all that. Just terribly shy. So he had friends in the sense that a very shy person has friends but not great intimacy or not with great attachment, partly because there were so many moves; …" [21]

In a letter to the Condons prior to Ronald's death, Natalie describes Ronald and herself as "casualties of the cold war".[133] Had it not been for Ronald's poor health and their reluctance to leave the United States for England, they could have weathered these storms and after a period of time, Ronald would have regained an academic foothold in British or American universities. Many others in similar situations, such as Condon, Urey, and Oppenheimer were subject to investigations, but were later proven to be loyal and reinstated to academic positions. There were legitimate security concerns at the time in the United States and UK and with Ronald and Natalie's wide network of acquaintances and past social activities, it can be understood how they became subjects of investigation. However, unless there is further evidence we are currently unaware of, it is unfortunate and unforgivable that despite numerous detailed investigations, which led to their clearance, the Gurneys could still not disentangle themselves from the web of accusations and suspicion.

As the early history of quantum mechanics and its applications and the sociopolitical atmosphere of that time are being revisited, we hope there will be greater interest in the extraordinary life, work, and writings of Ronald Wilfrid Gurney.

## Published papers, books, and reports by Ronald W. Gurney.

Gurney's affiliations stated in the publications are given in parentheses after the citation.

## Acknowledgments

The authors acknowledge the following individuals and universities who kindly provided or assisted in finding archival material.

Dr. Kathrine Condon; Prof. Stephen Fletcher, Loughborough University; Professor Daniel Sleator, Carnegie Mellon University; Nico Sleator; Dr. Brook Shilling, Reference Archivist, Sheridan Libraries Johns Hopkins University Archives; Executive Administration Team, School of Physics, University of Bristol; Elizabeth Wood, Corrine Mona, and Chip Calhoun, Librarians, AIP Niels Bohr Library and Archives; American Philosophical Society (E. U. Condon archives); Hannah Dale, Archives Manager, Cheltenham College; Catherine Uecker, Hanna Holborn Gray Special Collections Research Center, University of Chicago Library; Jennifer Donovan, Special Collections & Archives, UC San Diego Library; Alexandra Browne, College Archivist and Records Manager, Trinity Hall, Cambridge; Mutahara Mobashar, Manuscript Division, Library of Congress; and Vanessa Bismuth, Communications Manager, The Cavendish Laboratory, University of Cambridge.

## Disclosure statement

No potential conflict of interest was reported by the author(s).

## Funding

No funding received.

## Data Availability

Data sharing is not applicable to this article as no new data were created or analyzed in this study.

## ORCID

Brian Pollard http://orcid.org/0000-0001-9288-0954

Saman Alavi http://orcid.org/0000-0001-7334-6449

**Appendix**

Extracts of letters from Ronald Gurney's friends sent to Natalie Gurney after his death.

"A daring, speculative and creative thinker in things the Middle Ages never knew about, he was at the same time a wandering scholar such as the Middle Ages knew"

"It will be some comfort to remember that your husband has already achieved more immortality than most, even among men of his own stature, and that it is only in comparison with what could have been that one need sorry for science [*sic*]. This thought will be of no help in your sense of personal loss, but time will come when you can hold it with more comfort than now."

"I look away from this bleak time back to happier ones, and remember Ronald when he first came to America, a student scientist. I find a sort of comfort in the possibility that my son may be able to work some day in the field to which Ronald devoted himself. I wonder if other Americans who value Ronald might share my impulse to make some sort of contribution in his name, perhaps to aid another student scientist at Ronald's own university."

"Ronald's passing was a shock to me. Perhaps his being gone means something different to me than it would to most people. You see, ever since he gave his recommendation for my admission to college, I have always had an inward urge to prove myself equal to the trust he put in my native ability. Now that I cannot any longer show him the results of this inward desire to excel, I feel a sort of emptiness that one feels when he loses sight of a cherished goal. I did feel that if there was one man (besides my immediate family), for whom I would want to make a good showing in my profession, It was he. Now he will never know"

"I first met Ronald in July of 1933 at Cambridge, England, where he had gone for a few weeks during his summer vacation. He very kindly helped me with a theory I was working on with respect to the electrical potentials of glass membranes. A few years later when he wished to visit the United States, I secured a small stipend for him to lecture at Northwestern during the summer school. I remember vividly his lectures, his early morning running and his ability to play the piano and sing Japanese songs. His contributions to the scientific and social life of our Summer School at the time will never be forgotten by those of us who were here then."

"I know that Ronald had long lived in the shadow of death, and I had admired the resolution with which he and you had carried on, and with which he managed, despite all handicaps, to contribute to his research in a field to which he had made such notable contributions."

"As you know, I did not know him really well, and his condition. as well as the differences of our subjects prevented vivid communication. Yet I enjoyed his presence and his wit, and the critical twinkle in his eye. In short, I liked him."

"Ronald's loss is not only a personal one to those of us who knew him and respected him for his quiet worth, but a real loss for all who were indebted to his clear thinking and expression in abstruse areas of physics."

"It seems a terribly long time ago that he and I worked together on the Oxford book that we published in 1940, the book that never seemed worth working on because we all expected to be blown up. However we finished it mainly thanks to him… I very much look forward to seeing the new book when it comes out."

"His personality and his brilliant books made clear the great insight he was capable of as well as the gift of conveying his knowledge to others - a rare combination.

"Ronald was a brilliant scientist, I know, but he appealed to me because of his sensitivity and charm. I developed a special affection for him from the moment I first met him - a moment which might have been painful had it not been for the gentle humor with which he explained that he was the host and not a guest at his party".

"The memory of his gaiety as 'teacher' of the lovely English folk dances at Princeton when we were all younger and gayer can never be forgotten by any of us. Even during those years when he had much scholarly work to do, he gave us so generously of his leisure time and left us with the feeling of good times and of vigor and sweetness through his own delightful personality"

"His work as a physicist will be remembered long after all the rest of us are forgotten; but this is little consolation just now to those who have known and loved him as a person."

"Our friendship with Ronald was a genuinely precious and proud relationship. Now there is the better regret that we were unable to take advantage of it more during the years since we were all together at Aberdeen. However, his personality was unique, and was so rich that the memory of our friendship there will always be a source of satisfaction to us. We have the little cello-viola piece he wrote for us, which is charming …. And the dedication in Restoration style is in itself a masterpiece which has been proudly exhibited to any unsuspecting visitor whom we considered capable of appreciating it. Oh, I'm so very very glad we knew him."